\documentclass[12pt]{article}

\usepackage[margin = 1.5in]{geometry}
\usepackage{amsmath}
\usepackage{amsfonts}
\usepackage{bbm}
\usepackage{dsfont}
\usepackage{graphicx,psfrag,epsf}
\usepackage{color}
\usepackage{enumerate}
\usepackage{natbib}
\usepackage{hyperref}
\usepackage{caption}
\usepackage{subcaption}
\usepackage{float}
\usepackage{booktabs}
\usepackage{algorithm}
\usepackage{algpseudocode}

\allowdisplaybreaks

\newcommand{\blind}{1}

\newcommand{\V}{\mathcal{V}}

\newcommand{\rY}{Y^{(r)}}
\newcommand{\ry}{y^{(r)}}

\DeclareMathOperator*{\argmax}{arg\,max}

\newcommand\numberthis{\addtocounter{equation}{1}\tag{\theequation}}

\begin{document}

\def\spacingset#1{\renewcommand{\baselinestretch}%
{#1}\small\normalsize} \spacingset{1}

\if1\blind
{
  \title{\bf Maximum Likelihood Estimation for Network Models with Latent Geometry under Snowball Sampling}
  \author{Nurzhan Sapargali\\
    Department of Statistics, University of Munich\\
    and \\
    Sergio Buttazzo \\
    Department of Statistics, University of Munich \\
    and \\
    Göran Kauermann \\
    Department of Statistics, University of Munich}
  \maketitle
} \fi

\if0\blind
{
  \bigskip
  \bigskip
  \bigskip
  \begin{center}
    {\LARGE\bf Maximum Likelihood Estimation for Network Models with Latent Geometry under Snowball Sampling}
\end{center}
  \medskip
} \fi

\bigskip

\begin{abstract}
Snowball sampling is a widely used design for collecting network data from large or hard-to-reach populations, yet naive inference that ignores the sampling mechanism produces systematically biased parameter estimates.
We derive the exact likelihood of a multi-wave snowball sample for the class of continuous latent space (CLS) models, in which edges form independently conditional on latent vertex-level quantities, and show that conditional edge independence reduces the marginalization over unobserved network configurations to a closed-form expression valid across the entire CLS class.
We develop a stochastic Expectation-Maximization algorithm for the Euclidean latent distance model as a concrete implementation, and apply the framework to the large-scale co-inventor network of German semiconductor patent applicants by drawing multiple snowball samples.
We find that the naive procedure severely underestimates latent space variance, produces networks with nearly twice the observed edge count, and achieves a spectral goodness-of-fit nine times worse than the corrected model, which directly affects the quantitative interpretation of covariate effects.
\end{abstract}

\noindent
{\it Keywords:}  network sampling; model-based inference; sampling bias; latent space model
\vfill

\newpage
\spacingset{1.45}

\section{Introduction}
\label{sec:intro}
Large-scale networks are rarely fully observable.
Budgetary, logistical, and ethical constraints routinely prevent researchers from recording every tie in a population network, and even when full observation is feasible, the network may simply be too large for likelihood-based inference to remain computationally tractable.
Many canonical statistical network models such as exponential random graph models (ERGMs) \citep{frank1986markov, robins2007intro} or latent position models \citep{kaur2023latent} scale poorly with network size.
Whether through intractable normalizing constants or Markov chain Monte Carlo (MCMC) updates that scale quadratically with network size, likelihood evaluations or posterior sampling render the exact inference on fully observed large networks computationally prohibitive without further approximation.
In both settings, sampling from the network emerges as a natural and practical alternative.
By working with a tractable sub-network one can draw inferences about the full population graph.
However, the statistical validity of this approach depends critically on whether the sampling mechanism is properly accounted for in the inferential procedure.

Among the many network sampling designs that have been considered \citep{frank1977survey, hu2013survey, zhang2017graph}, snowball sampling stands out as a natural, practical and widely used method for collecting network data.
Introduced formally by \citet{goodman1961snowball}, the sampling scheme begins with a small set of seed vertices and expands by recruiting network neighbors of the currently sampled set, wave by wave, until a stopping criterion is met.
With a long history in sociology, epidemiology, and the study of online networks, its adaptive, contact-tracing structure makes it a natural fit for hard-to-reach populations \citep{heckathorn2017network}, while its breadth-first traversal of the network means that a modest number of waves can cover a substantial fraction of a large sparse graph at low cost.

The same adaptive structure is, however, the source of its principal inferential difficulty.
Because vertex inclusion in a snowball sample depends on network connectivity, the sample is far from a representative subgraph.
Vertices with high degree have a systematically higher probability of being recruited, and dense regions of the network are over-represented relative to sparse peripheries \citep{johnson1989estimating, snijders1992estimation, illenberger2012estimating}.
The observed subgraph therefore looks more clustered and homogeneous than the underlying population network, so a practitioner who fits a network model to the snowball sample naively, i.e. as if it were the full population network, will systematically underestimate the heterogeneity and variance of the population parameters.
This bias is not a mere technicality: as our application and simulation demonstrate, it can alter conclusions in a quantitatively dramatic fashion.

The literature on inference from snowball samples spans several decades and methodological traditions.
The earliest contributions are design-based:
\citet{goodman1961snowball}, \citet{frank1977survey}, \citet{snijders1992estimation}, and \citet{frank1994estimating} derive estimators for scalar network parameters such as edge density, degree moments, population size by exploiting the known inclusion probabilities of the design, without assuming a parametric model for the population network.
More recent design-based work, including \citet{illenberger2012estimating} and \citet{oguzalper2023snowball}, extends this tradition to a broader range of graph statistics and provides a rigorous general framework for Horvitz--Thompson estimation under multi-wave snowball sampling \citep{zhang2017graph}.
These methods impose minimal assumptions, but they are targeted at specific estimands and do not yield a full likelihood for the network model.

Model-based approaches, by contrast, seek to estimate the parameters of a generative model for the population network from the observed sample.
\citet{handcock2010modeling} provide the foundational framework:
Following \citet{thompson2000model}, they define a network sampling design as \emph{amenable} to a given model if the sampling mechanism is ignorable for likelihood-based inference, and show that the resulting likelihood can be expressed as a marginalization of the full-network likelihood over all population networks consistent with the observed sample.
This is a principled and general formulation, but the marginalization involves a sum over exponentially many network configurations, compounding the already-demanding computational problem of normalizing constant evaluation.
\citet{pattison2013conditional} propose a conditional maximum likelihood approach for the ERGM that sidesteps this difficulty by conditioning on the sufficient statistics of the unobserved part of the network, enabling practical estimation from snowball samples.
\citet{stivala2016snowball} extend this to large networks via parallel computation and a meta-analytic combination of estimates across samples.
\citet{vincent2022estimating} develop a Bayesian data augmentation procedure under a generalized stochastic block model, deriving an observed likelihood by summing over unobserved stratum assignments and adjacency values.
The resulting sum is again analytically cumbersome, leading the authors to rely on Gibbs sampling rather than direct likelihood evaluation.

Each of these contributions share a common limitation: they are either model-specific, or they reduce to marginalizing over the unobserved network configurations, which is a computationally expensive operation that must be re-engineered for each new model class.
Absent from the literature is an exact, closed-form likelihood for the snowball sample that is valid across a broad class of network models and that can be evaluated efficiently without enumerating unobserved networks.

This paper fills that gap.
The key observation is that the distribution of the snowball sample factorizes in a particularly transparent way for any network model in which edges form independently across vertex pairs conditional on latent vertex-level quantities.
This class of models, termed \emph{continuous latent space} (CLS) models, following \citet{smith2019geometry}, is broad enough to cover many of the most widely used and well-studied network models, including latent space models, random dot product graphs \citep{young2007random}, graphon models \citep[e.g.][]{choi2014co, gao2015rate, sischka2025stochastic}, and stochastic block models \citep{holland1983stochastic, snijders1997estimation}.
Conditional edge independence implies that the marginalization over unobserved network configurations, which is intractable in general, reduces to a simple product over vertex pairs, where the likelihood of the snowball sample is a closed-form expression in the model parameters and the observed wave sets.
The result resolves, for the CLS class, the open problem identified by \citet{crane2018probabilistic} of computing the distribution of multi-wave snowball samples from a general population network model.

Building on the closed-form likelihood \citet{sapargali2026exact} establish for the Erd\H{o}s--R\'{e}nyi model, whose unconditional edge independence yields particularly clean expressions and provides the intuition for the general case, we extend the result to the full CLS class.
As a concrete implementation of the resulting likelihood, we develop a stochastic Expectation-Maximization (sEM) algorithm for the distance model of \citet{hoff2002latent}.
Crucially, the correction integrates naturally with existing MCMC samplers for latent position inference, adding only a single low-dimensional integral, which reflects the expected exclusion probability of an unobserved vertex and whose efficient approximation via quasi-Monte Carlo methods keeps the computational overhead modest.

We demonstrate the methodology on the co-inventor network of German electrical engineering patent applicants \citep{fritz2023modelling}, a network of approximately $6{,}000$ inventors and $21{,}000$ co-invention ties exhibiting pronounced small-world characteristics, high clustering and low average shortest path length, as well as heavy-tailed degree distributions.
Simple latent space models based on Euclidean geometry are known to perform poorly on such networks, as \citet{rastelli2016properties} show that the latent position model in its original form cannot represent small-world behaviour, nor heavy-tailed degree distributions as network size grows.
Yet this incompatibility has, to our knowledge, never been examined empirically at scale, in part because MCMC-based inference over latent positions has historically been prohibitive for networks of this size.
Our framework thus simultaneously enables the first empirical stress test of the distance model on a large-scale network exhibiting genuine small-world properties and provides a direct comparison between the corrected and naive procedures.

Our results show that even in this challenging setting, the distance model can recover meaningful latent structure, much more than the naive approach suggests.
In a large-scale study drawing $500$ independent snowball samples and aggregating estimates via a robust mode-based meta-analytic estimator \citep{hartwig2020median}, we find that the naive approach severely underestimates the spread of the latent space, producing networks with nearly twice the observed edge count, and achieving a spectral goodness-of-fit \citep{shore2015spectral} nine times worse than the corrected model.
The corrected procedure, by contrast, recovers a latent space variance roughly eight times larger and better reproduces the degree distribution and spectral properties of the observed network.
While both procedures yield qualitatively similar patterns of gender homophily and geographic proximity to \citet{fritz2023modelling}, the corrected model produces substantially lower and more variable edge probabilities as a direct consequence of the larger estimated latent space variance, which materially affects the quantitative interpretation of the results.

The remainder of the paper is organized as follows.
Section~\ref{sec:likelihood} develops the likelihood of a snowball sample.
After fixing notation and the sampling scheme (Section~\ref{sec:notation}), we introduce the continuous latent space (CLS) model class (Section~\ref{subsec:cls_model}) and recap the exact likelihood for the Erd\H{o}s--R\'{e}nyi case established by \citet{sapargali2026exact} as its simplest member (Section~\ref{subsec:er_recap}).
Section~\ref{subsec:cls_extension} extends this result to the general CLS class, establishes identifiability of the latent positions of unsampled vertices, and discusses non-ignorability of ego selection.
Section~\ref{sec:distance} focuses on the latent distance model, develops the sEM algorithm, and presents the simulation study.
Section~\ref{sec:application} presents the patent co-inventor application.
Section~\ref{sec:discussion} concludes with a discussion of limitations and directions for future work.

\section{Likelihood of a Snowball Sample}
\label{sec:likelihood}

In this section, we develop the likelihood of a snowball sample, building up from basic notation to the general CLS class.
After fixing notation for the population network and the sampling scheme, we introduce the CLS model class and its population distribution, recap the Erd\H{o}s--R\'{e}nyi case as its simplest member, and extend the resulting likelihood to the general CLS class.

\subsection{Notation and Sampling Scheme}
\label{sec:notation}

Let $G = (V, E)$ denote the population network, an undirected graph on vertex set $V = \{1, \dots, N\}$ with edge set $E$.
For a finite set $A$, let $A^2 = \{\{i,j\} : i,j \in A,\, i \neq j\}$ denote the collection of unordered pairs of distinct elements of $A$, so that $E \subseteq V^2$ rules out self-loops and multiple edges between any pair of vertices.

We encode $G$ through its adjacency matrix $Y = (Y_{i,j})_{1 \leq i,j \leq N}$, with $Y_{i,j} = Y_{j,i} = 1$ whenever $\{i,j\} \in E$ and zero otherwise.
Since the population network is itself treated as random, $Y$ is a random matrix, and we write $y$ for its realized value.

An $r$-wave snowball sample from $G$ is built up wave by wave.
The initial wave $\V^{(0)} = \{v_0\}$ contains a single vertex $v_0 \in V$, the \textsl{ego}, either fixed in advance or drawn according to some sampling scheme.
When $v_0$ is random, everything that follows is conditioned on its observed value.

A vertex enters the first wave $\V^{(1)}$ if it is a neighbor of the ego:
\begin{equation*}
    \V^{(1)} = \{j \in V : \{v_0, j\} \in E\}.
\end{equation*}
Subsequent waves recruit further outward.
For $k = 2, \dots, r$, the $k$-th wave $\V^{(k)}$ collects any neighbor of a vertex in $\V^{(k-1)}$ not already present in an earlier wave:
\begin{equation*}
    \V^{(k)} = \Bigl(\bigcup_{i \in \V^{(k-1)}} \{j \in V : \{i,j\} \in E\} \Bigr) \mathbin{\big\backslash} \Bigl(\bigcup_{s=0}^{k-1} \V^{(s)}\Bigr).
\end{equation*}

\begin{figure}[!ht]
    \centering
    \includegraphics[scale = 0.5]{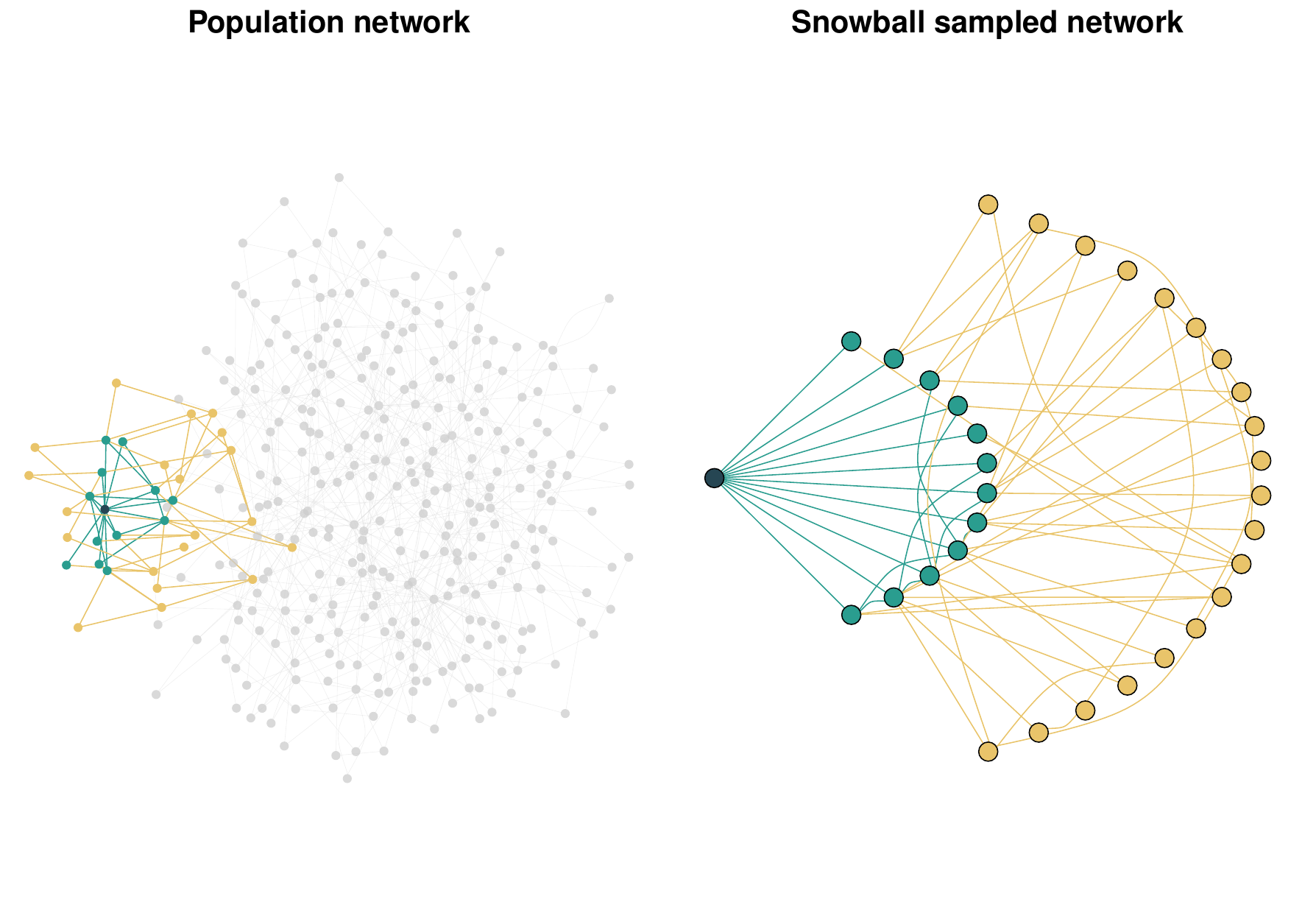}
    \caption{Graphical illustration of a 2-wave snowball sample.}
    \label{fig:pa_graph}
\end{figure}

The sampled graph $G^{(r)} = (\bigcup_{k=0}^r \V^{(k)}, E^{(r)})$ consists of the sampled vertex set $\bigcup_{k=0}^r \V^{(k)}$ and an edge set $E^{(r)} \subseteq \bigl(\bigcup_{k=0}^r \V^{(k)}\bigr)^2$, containing all edges among the sampled vertices.
The corresponding sampled adjacency matrix $\rY$ is defined analogously to $Y$, restricted to pairs $\{i,j\} \in \bigl(\bigcup_{k=0}^r \V^{(k)}\bigr)^2$: $\rY_{i,j} = \rY_{j,i} = 1$ if $\{i,j\} \in E^{(r)}$, and zero otherwise.
Figure~\ref{fig:pa_graph} shows a 2-wave sample from a population network, with wave membership indicated by color.

Because wave composition depends on the unknown structure of the population network, the wave sets $\V^{(0)}, \V^{(1)}, \dots, \V^{(r)}$ are random along with $\rY$, and it is this joint randomness of which vertices are observed at all, and which edges are observed among them that complicates inference from a snowball sample.

\subsection{The Continuous Latent Space Model Class}
\label{subsec:cls_model}

To accommodate well-documented empirical features of real-world networks, such as transitivity and heavy-tailed degree distributions \citep{newman2018networks}, we work with a general class of network models introduced by \citet{smith2019geometry}, in which the observed network is generated through an unobserved, latent lower-dimensional space.
Each vertex is assigned a latent position in this space, and edges form independently across vertex pairs conditional on these positions, with edge probabilities determined by a similarity function of the corresponding latent coordinates.
This yields a flexible class of models in which dependence between edges is governed by the geometry of the latent space, thereby capturing complex network features such as transitivity and clustering.
\citet{smith2019geometry} term this class continuous latent space (CLS) models.

For undirected networks, the general form of a CLS model is:
\begin{gather*}
    Y_{i,j} \mid Z_i = z_i, Z_j = z_j \overset{ind}{\sim}\text{Bernoulli}(\pi_\alpha(z_i, z_j)) \quad \text{for all} \quad \{i,j\} \in V^2 \\
    \text{logit}(\pi_\alpha(z_i, z_j)) = \alpha - s(z_i, z_j) \\
    Z_i \in \mathcal{Z}^d, \quad Z_i \overset{ind}{\sim} F_\psi(z_i) \quad \text{for all} \quad i \in V,\numberthis \label{eq:latgeom_model}
\end{gather*}
where $Z_i$ is the latent position of vertex $i$ in a $d$-dimensional latent space $\mathcal{Z}^d$, $F_\psi(z)$ is a distribution over the latent space parameterized by $\psi$, and $\alpha \in \mathbb{R}$ is an intercept controlling the overall sparsity of the network.
The similarity function $s: \mathcal{Z}^d \times \mathcal{Z}^d \to \mathbb{R}$ governs the probability of an edge between vertices $i$ and $j$ through their latent positions $z_i$ and $z_j$, with the specific choice of $s$ encoding the structural assumptions of the model, e.g., whether vertices tend to connect to geometrically proximate others, to those in the same latent community, or to those with complementary latent profiles.

Many popular models in the literature arise as special cases of \eqref{eq:latgeom_model} through specific choices of $s(z_i, z_j)$, $\mathcal{Z}^d$, and $F_\psi(z)$.
The distance model of \citet{hoff2002latent} is obtained by setting $F_\psi(z)$ to be a multivariate normal distribution over $\mathcal{Z}^d = \mathbb{R}^d$ and $s(z_i, z_j) = \|z_i - z_j\|$, where $\|\cdot\|$ denotes the Euclidean distance.
Hyperbolic latent space models \citep{krioukov2010hyperbolic} similarly use a distance-based similarity function, but embed vertices in a negatively curved hyperbolic space, which naturally accommodates the heavy-tailed degree distributions and hierarchical structure observed in many real-world networks.
As we recap in Section~\ref{subsec:er_recap}, the Erd\H{o}s--R\'{e}nyi model is likewise nested in \eqref{eq:latgeom_model}, as the degenerate case in which $\pi_\alpha(z_i, z_j)$ does not depend on $z_i, z_j$ at all.

Several other well-known network models are closely related to \eqref{eq:latgeom_model}, though they differ in their choice of link function.
These include random dot product graph \citep{young2007random}, graphon \citep{lovasz2006limits}, and stochastic block models \citep{holland1983stochastic, snijders1997estimation}, which model edge probabilities directly rather than through a logit link.
Despite this difference, the structural insights derived from \eqref{eq:latgeom_model} are directly relevant to this wider class of models.

Given the flexibility and expressiveness of \eqref{eq:latgeom_model}, we adopt it as the population model for networks from which snowball samples are drawn.
We note that \eqref{eq:latgeom_model} can be extended to include covariates, either as part of the similarity function $s(z_i, z_j)$ or as an additional term in the linear predictor.
For clarity, we focus on the covariate-free case, which already reveals the key inferential challenges introduced by snowball sampling in latent geometry models, without the additional complication of missing covariates for unobserved vertices.
The inferential goal is to estimate the model parameters $\alpha$ and $\psi$, together with the latent positions $Z_1, \dots, Z_N$, using only the observed snowball sample.

\subsection{The Erd\H{o}s--R\'{e}nyi Case}
\label{subsec:er_recap}

The Erd\H{o}s--R\'{e}nyi (ER) model arises from \eqref{eq:latgeom_model} in the degenerate case where the edge probability $\pi_\alpha(z_i, z_j) \equiv \pi$ does not depend on the latent positions $z_i, z_j$ at all, so that
\begin{equation*}
    Y_{i,j} \sim \text{Bernoulli}(\pi) \quad \text{for all} \quad \{i,j\} \in V^2,
\end{equation*}
independently across vertex pairs, where $0 < \pi < 1$ is the unknown edge probability.
As the simplest member of the CLS class, the ER model offers a transparent illustration of how the snowball sampling design enters the likelihood before we turn to the general case in Section~\ref{subsec:cls_extension}.
The full derivation for this case is given by \citet{sapargali2026exact}.
We recap the results relevant to the general CLS derivation here.

Suppose an $r$-wave snowball sample with wave sets $\V^{(0)}, \V^{(1)}, \dots, \V^{(r)}$ and induced adjacency matrix $\rY$ is drawn from an ER population network with edge probability $\pi$.
Because edges are unconditionally independent under the ER model, a vertex enters wave $k$ if and only if it has at least one edge to wave $k-1$, and, conditional on the wave sets $\V^{(0)}, \dots, \V^{(k-1)}$, this event is independent across the vertices not yet sampled.

To express the resulting likelihood, we partition the sampled vertex pairs $\bigl(\bigcup_{k=0}^r \V^{(k)}\bigr)^2$ according to wave membership into the sets of vertex pairs within the same wave, between adjacent waves, and between non-adjacent waves, respectively:
\begin{align*}
    W &= \Bigl\{ \{i,j\} \in \bigl(\textstyle\bigcup_{k=0}^r \V^{(k)}\bigr)^2 : i, j \in \V^{(t)}, \ t = 1, \dots, r \Bigr\}, \\
    A &= \Bigl\{ \{i,j\} \in \bigl(\textstyle\bigcup_{k=0}^r \V^{(k)}\bigr)^2 : i \in \V^{(t)}, \ j \in \V^{(t-1)}, \ t = 1, \dots, r \Bigr\}, \\
    J &= \bigl(\textstyle\bigcup_{k=0}^r \V^{(k)}\bigr)^2 \setminus (W \cup A),
\end{align*}
By construction $\ry_{i,j} = 0$ for all $\{i,j\} \in J$: an edge between non-adjacent waves would have altered the wave-recruitment process itself, so no such edge can appear in a valid snowball sample.

Given the wave sets, the ER model's edge independence lets the joint probability of the sampled wave sets and adjacency matrix factorize directly, but not every configuration of $\ry$ is consistent with having observed those wave sets in the first place.
Two constraints, inherent to the recruitment process itself, restrict the support of $\rY$ conditional on the realizations of $\V^{(0)}, \dots, \V^{(r)}$: (1) no edges connect non-adjacent waves, since such an edge would have drawn the more distant vertex into an earlier wave than observed; and (2) every vertex in wave $k \geq 1$ has at least one edge to wave $k-1$, as this is precisely what recruited it.

Let $\mathcal{Y}$ be the set of adjacency matrices on $\bigcup_{k=0}^r V^{(k)}$ satisfying both constraints given the observed waves.
\citet{sapargali2026exact} show that, conditional on the ego, the joint probability of the sampled wave sets and adjacency matrix is
\begin{align*}
    P\bigl(\rY = \ry, \V^{(1)} = V^{(1)}, \dots, \V^{(r)} = V^{(r)} \mid \V^{(0)} = V^{(0)}\bigr) &= \left[\prod_{\{i,j\} \in W \cup A \cup J} \pi^{\ry_{i,j}}(1 - \pi)^{1 - \ry_{i,j}}\right] \\
    &\quad \cdot (1 - \pi)^{|\mathcal{U}| \sum_{k = 0}^{r - 1} n_k} \, \mathds{1}\{\ry \in \mathcal{Y}\}, \numberthis \label{eq:er_joint_prob_complete}
\end{align*}
where $\mathcal{U} = V \setminus \bigcup_{k=0}^r V^{(k)}$ is the set of unsampled vertices, $V^{(1)}, \dots, V^{(r)}$ are realizations of $\V^{(1)}, \dots, \V^{(r)}$ and $n_k = |V^{(k)}|$.
The first factor in \eqref{eq:er_joint_prob_complete} is simply the ER likelihood of the observed edges and non-edges within the sample.
The second factor is the probability that every unsampled vertex has no edge to any vertex in waves $0$ through $r-1$ under the ER model, which is precisely the event that must occur for those vertices to remain unsampled.
It is this second factor, entirely absent from a naive treatment of the sample as a complete ER graph, that carries the information the design contributes about $\pi$.
The indicator confines the expression to the wave-consistent configurations captured by $\mathcal{Y}$.

For the remainder of this paper, we write probabilities as functions of the observed realizations only for notational simplicity, e.g., writing the left-hand side of \eqref{eq:er_joint_prob_complete} as $P(\ry, V^{(1)}, \dots, V^{(r)} \mid V^{(0)})$.

Writing $n = (n_1, \dots, n_r)$ for the wave sizes, \citet{sapargali2026exact} show that \eqref{eq:er_joint_prob_complete} can equivalently be written as
\begin{equation*}
    P\bigl(\ry, V^{(1)}, \dots, V^{(r)} \mid V^{(0)}\bigr) = \exp\Bigl\{ |E^{(r)}| \log \tfrac{\pi}{1-\pi} + T(n) \log(1-\pi) \Bigr\} \mathds{1}\{\ry \in \mathcal{Y}\},
\end{equation*}
where $|E^{(r)}| = \sum_{\{i,j\} \in W \cup A} \ry_{i,j}$ is the number of observed edges and $T(n) = |W| + |A| + |J| + |\mathcal{U}|\sum_{k=0}^{r-1} n_k$.
This identifies $\bigl(|E^{(r)}|, T(n)\bigr)$ as the minimal sufficient statistic of a curved exponential family in $\pi$, and yields the maximum likelihood estimator
\begin{equation*}
    \hat{\pi} = \frac{|E^{(r)}|}{T(n)},
\end{equation*}
which corrects the upward bias of the naive estimator obtained by treating the sample as a complete ER graph on $\bigcup_{k=0}^r V^{(k)}$.

\subsection{The General CLS Case}
\label{subsec:cls_extension}

The derivation of \eqref{eq:er_joint_prob_complete} relies only on edge independence of the ER model, a property the general CLS model retains conditional on the latent positions $Z$.
Accounting for the pair-specific edge probability $\pi_\alpha(z_i, z_j)$ in place of the constant $\pi$, the results of Section~\ref{subsec:er_recap} thus extend directly.

Let $Z = (Z_1, \dots, Z_N)^\top \in \mathcal{Z}^{N \times d}$ denote the matrix of latent positions for all vertices in the population, with realization $z = (z_1, \dots, z_N)^\top$.
Mirroring \eqref{eq:er_joint_prob_complete}, the joint probability of the sampled wave sets and adjacency matrix under the CLS model, conditional on $V^{(0)}$ and $Z = z$, is
\begin{align}
    P\bigl(\ry, V^{(1)}, \dots, V^{(r)} \mid V^{(0)}, z\bigr) &= \left[\prod_{\{i,j\} \in W \cup A \cup J} \pi_\alpha(z_i, z_j)^{\ry_{i,j}} \bigl(1 - \pi_\alpha(z_i, z_j)\bigr)^{1 - \ry_{i,j}}\right] \notag \\
    &\quad \cdot \left[\prod_{i \in \mathcal{U}} \prod_{j \in \bigcup_{s=0}^{r-1} V^{(s)}} \bigl(1 - \pi_\alpha(z_i, z_j)\bigr)\right] \mathds{1}\{\ry \in \mathcal{Y}\}, \label{eq:cls_joint_prob}
\end{align}
where $\mathcal{U}$, $\mathcal{Y}$, $W$, $A$, and $J$ retain their meaning from Section~\ref{subsec:er_recap}.
As in the ER case, the first factor is the CLS likelihood of the observed edges and non-edges within the sample.
The second factor is now a product over the individual pairs linking unsampled vertices to waves $0$ through $r-1$, rather than a single power of a common probability, as $\pi_\alpha(z_i, z_j)$ varies with the latent positions of both vertices.
The full derivation of \eqref{eq:cls_joint_prob} is given in Appendix~\ref{app:cls_likelihood_derivation}.

Examining \eqref{eq:cls_joint_prob} directly, the latent positions of unsampled vertices enter only through the second factor, $\prod_{i \in \mathcal{U}} \prod_{j \in \bigcup_{s=0}^{r-1} V^{(s)}} (1 - \pi_\alpha(z_i, z_j))$, which is the joint exclusion probability, i.e. the probability that no unsampled vertex connects to any vertex in waves $0$ through $r-1$.
Consequently, the latent positions of unsampled vertices are identifiable only up to the equivalence class of configurations that yield the same joint exclusion probability.
Appendix~\ref{app:lsm_counterexample} constructs a worked counterexample using the distance model of \citet{hoff2002latent}, in which the positions of unsampled vertices can be rescaled without altering the likelihood.

The dependence of the sample probability on the latent positions $Z = z$ of the entire population also lets us formally demonstrate a second, previously known problem of non-ignorable ego selection.
Letting $\theta = (\alpha, \psi)$ denote the vector of population parameters, consider the joint density of the sampled adjacency matrix, wave sets, and latent positions of all vertices, conditional on the ego vertex $V^{(0)}$:
\begin{equation} \label{eq:cls_full_likelihood}
    f_\theta(\ry, V^{(1)}, \dots, V^{(r)}, z \mid V^{(0)}) =
    P(\ry, V^{(1)}, \dots, V^{(r)} \mid V^{(0)}, z)\, f(z \mid V^{(0)}),
\end{equation}
where $f(z \mid V^{(0)})$ is the posterior density of the latent positions given the choice of ego vertex.
By Bayes' theorem,
\begin{equation*}
    f(z \mid V^{(0)}) = \frac{P(V^{(0)} \mid z)\, f_\psi(z)}{P(V^{(0)})},
\end{equation*}
where $f_\psi(z)$ is the population prior density of the latent positions.

Unless the ego selection mechanism $P(V^{(0)} \mid z)$ is independent of the latent positions $z$, the density $f(z \mid V^{(0)})$ differs from the population prior $f_\psi(z)$, and the likelihood \eqref{eq:cls_full_likelihood} depends on the choice of ego vertex, rendering the ego selection mechanism non-ignorable.
The independence condition holds, for example, when the ego is selected uniformly at random, based on an observed covariate independent of the latent positions, or by a deterministic mechanism unrelated to network features.
It fails when selection is based on network features such as degree or centrality, which are functions of the latent positions.

To see that $f(z \mid V^{(0)})$ can genuinely differ from $f_\psi(z)$, consider degree-proportional selection, where vertex $i$ is chosen with probability proportional to its degree.
Computing $P(V^{(0)} \mid z)$ with $V^{(0)} = \{i\}$ requires marginalizing over the joint distribution of the degree $D_i = \sum_{j \neq i} Y_{i,j}$ and the residual degree sum $K = \sum_{l \neq i} \sum_{j \neq l, j \neq i} Y_{l,j}$:
\begin{align*}
    P(V^{(0)} \mid z) &= \sum_{k = 0}^{(N - 1)(N - 2)} \sum_{d = 0}^{N - 1} P(\V^{(0)} = V^{(0)}, D_i = d, K = k \mid z) \\
    &= \sum_{k = 0}^{(N - 1)(N - 2)} \sum_{d = 0}^{N - 1} P(\V^{(0)} = V^{(0)} \mid D_i = d, K = k, z) P(D_i = d \mid K = k, z)P(K = k \mid z) \\
    &= \sum_{k = 0}^{(N - 1)(N - 2)} \sum_{d = 0}^{N - 1} \frac{d}{k + d} P(D_i = d \mid z) P(K = k \mid z),
\end{align*}
where $P(D_i = d \mid z)$ and $P(K = k \mid z)$ are Poisson-binomial distributions with parameters determined by the latent positions $z$, and the factorization in the last line follows from the conditional independence of $D_i$ and $K$ given $z$.
Even in this simple case, $P(V^{(0)} \mid z)$ is a nontrivial function of $z$, so $f(z \mid V^{(0)})$ departs from $f_\psi(z)$.

The difficulty of modelling the ego selection mechanism is not unique to CLS models.
To our knowledge, no existing model-based approach to snowball sampling addresses non-ignorable ego selection.
In the ERGM literature, \citet{pattison2013conditional} assume that the ego set is chosen independently of the network, while \citet{vincent2022estimating} in their treatment of the generalized stochastic block model acknowledge that degree-proportional ego selection induces systematic bias but leave its treatment to future work.
While there are many possibilities for the ego selection mechanism, modelling it explicitly is not the primary focus of this paper.
The survey of the snowball sampling literature presented by \citet{illenberger2012estimating} finds uniform seed selection to be the predominant design convention, with degree-based recruitment schemes such as respondent-driven sampling a notable exception.
We therefore proceed under the assumption of ignorable ego selection, $f(z \mid V^{(0)}) = f_\psi(z)$, which allows us to develop a practical estimation procedure without the added complexity of modelling the ego selection mechanism explicitly.

\section{Example: Distance Model}
\label{sec:distance}
In this section, we return to the distance model of \citet{hoff2002latent}, introduced in Section~\ref{subsec:cls_model} as a special case of the CLS class, and apply the general likelihood results of Section~\ref{sec:likelihood} to this concrete specification.
This model is widely used in the literature and captures key features of real-world networks, such as transitivity and clustering, while maintaining a structure that admits tractable inference \citep{kaur2023latent}.
The results extend readily to other CLS model variants, as the key inferential challenges introduced by snowball sampling are shared across different specifications of the similarity function and the latent position distribution.
Recalling its specification:
\begin{gather*}
    Y_{i,j} \mid Z_i = z_i, Z_j = z_j \overset{ind}{\sim} \text{Bernoulli}(\pi_\alpha(z_i, z_j)) \quad \text{for all} \quad \{i,j\} \in V^2 \\
    \text{logit}(\pi_\alpha(z_i, z_j)) = \alpha - \|z_i - z_j\| \\
    Z_i \in \mathbb{R}^d, \quad Z_i \overset{iid}{\sim} N(0, \psi I_d) \quad \text{for all} \quad i \in V, \numberthis \label{eq:hoff_model}
\end{gather*}
where $\alpha \in \mathbb{R}$ is an intercept controlling the overall sparsity
of the network, $\psi > 0$ is the variance of the latent positions, and $I_d$
is the $d$-dimensional identity matrix.

\subsection{Parameter Estimation}
\label{subsec:lsm_estimation}

To address the identifiability issues established in Section~\ref{subsec:cls_extension}, we proceed in three steps.
We integrate out the latent positions of unsampled vertices from the joint likelihood, treat the positions of sampled vertices as missing data, and estimate the population parameters $\theta = (\alpha, \psi)$ via the stochastic Expectation-Maximization (sEM) algorithm \citep{celeux1985sem, diebolt1995stochastic}.

Following the assumption of ignorable ego selection, the joint density of the sampled data and the latent positions of all vertices is:
\begin{align*}
    f_\theta(\ry, V^{(1)}, \dots, V^{(r)}, z \mid V^{(0)}) =& \ P(\ry, V^{(1)}, \dots, V^{(r)} \mid V^{(0)}, z)\, f_\psi(z) \\
    =& \ \prod_{\{i,j\} \in A \cup W \cup J} \pi_\alpha(z_i, z_j)^{\ry_{i,j}} (1 - \pi_\alpha(z_i,z_j))^{1 - \ry_{i,j}} \\
    &\cdot \prod_{i \in \mathcal{U}} \prod_{j \in \bigcup_{s = 0}^{r - 1} V^{(s)}} (1 - \pi_\alpha(z_i,z_j)) \cdot \prod_{i \in V} f_\psi(z_i).
\end{align*}
Integrating out the latent positions of unsampled vertices yields the marginal likelihood of the observed data and the latent positions of the sampled vertices:
\begin{align*}
    f_\theta(\ry, V^{(1)}, \dots, V^{(r)}, z_{V\setminus\mathcal{U}} \mid V^{(0)}) =& \prod_{\{i,j\} \in A \cup W \cup J} \pi_\alpha(z_i, z_j)^{\ry_{i,j}} (1 - \pi_\alpha(z_i,z_j))^{1 - \ry_{i,j}} \prod_{i \in V \setminus \mathcal{U}} f_\psi(z_i) \\
    &\cdot\mathbb{E}_{Z_i}\Big[\prod_{j \in \bigcup_{s = 0}^{r - 1} V^{(s)}}(1 - \pi_\alpha(Z_i, z_j)) \Big]^{|\mathcal{U}|}. \numberthis \label{eq:complete_data_like}
\end{align*}
Here $z_{V\setminus\mathcal{U}}$ denotes the latent positions of the sampled vertices, and
\begin{equation} \label{eq:lsm_exclusion_prob}
    \mathbb{E}_{Z_i}\Big[\prod_{j \in \bigcup_{s = 0}^{r - 1} V^{(s)}}(1 - \pi_\alpha(Z_i, z_j)) \Big] = \int_{\mathcal{Z}^d} f_\psi(t) \prod_{j \in \bigcup_{s = 0}^{r - 1} V^{(s)}}(1 - \pi_\alpha(t,z_j))\, dt
\end{equation}
is the expected exclusion probability of a vertex with latent position $Z_i$ under the chosen prior distribution $f_\psi(z)$.

Beyond resolving the identifiability issues of Section~\ref{subsec:cls_extension}, integrating out the unsampled positions yields a substantial computational advantage.
Rather than performing inference on the latent positions of all $N$ vertices, inference is reduced to the $N - |\mathcal{U}|$ sampled vertices only, a significant reduction in networks where $|\mathcal{U}|$ is large.
The contribution of the unsampled vertices is absorbed entirely into the expected exclusion probability \eqref{eq:lsm_exclusion_prob}, which is a single low-dimensional integral that, while intractable in closed form, can be approximated efficiently at a computational cost far lower than explicit inference over $|\mathcal{U}|$ individual positions.

The corresponding observed data log-likelihood is:
\begin{equation*}
    \log L(\theta) = \log \int_{\mathcal{Z}^{d \sum_{k = 0}^r n_k}} f_\theta(\ry, V^{(1)}, \dots, V^{(r)}, z_{V\setminus\mathcal{U}} \mid V^{(0)})\, dz_{V\setminus\mathcal{U}},
\end{equation*}
where $n_k = |V^{(k)}|$ denotes the size of the $k$-th wave.

Direct maximization of $\log L(\theta)$ is intractable due to the high-dimensional integral over the latent positions of sampled vertices.
We therefore treat $z_{V\setminus\mathcal{U}}$ as missing data and apply the sEM algorithm, starting from an initial estimate $\hat{\theta}^{(0)} = (\hat{\alpha}^{(0)}, \hat{\psi}^{(0)})$ and iterating the following two steps until convergence:

\begin{description}
    \item[Stochastic E-step.] Given the current parameter estimates $\hat{\theta}^{(t)}$, draw a sample $z_{V\setminus\mathcal{U}}^{(t)}$ from the posterior distribution $f_{\hat{\theta}^{(t)}}(z_{V\setminus\mathcal{U}} \mid \ry, V^{(0)}, \dots, V^{(r)})$ via a short Markov Chain Monte Carlo (MCMC) run.
    \item[M-step.] Update the parameter estimates by maximizing the complete data log-likelihood \eqref{eq:complete_data_like} at the sampled latent positions:
    \begin{equation*}
        \hat{\theta}^{(t+1)} = \argmax_{\theta}\ \log f_\theta\!\left(\ry, V^{(1)}, \dots, V^{(r)}, z_{V\setminus\mathcal{U}}^{(t)} \mid V^{(0)}\right).
    \end{equation*}
\end{description}

The stochastic E-step requires sampling from the posterior distribution of the latent positions of the sampled vertices.
The standard approach for distance models is a Metropolis-within-Gibbs sampler, which perfroms a Gibbs update for each latent position $Z_i$ sequentially via a Metropolis-Hastings step with a Gaussian random walk proposal \citep{kaur2023latent}.
While the Metropolis-within-Gibbs sampler is straightforward to implement in our setting, we leverage the Gaussian prior of the distance model and replace the Metropolis-Hastings step with the Elliptical Slice Sampler (ESS) of \citet{murray2010elliptical} for a tuning-free update of each $Z_i$.
The reduction from $N$ to $N - |\mathcal{U}|$ positions translates directly into a proportional reduction in the number of Gibbs updates per iteration, which is the primary computational benefit of snowball sampling in this setting.
The full conditional distribution of $Z_i$ at iteration $t$ is proportional to:
\begin{align*}
    f_{\hat{\theta}^{(t)}}(z_i \mid \ry, V^{(0)}, \dots, V^{(r)}, z_{-i}) \propto& \ f_{\hat{\psi}^{(t)}}(z_i) \prod_{j  \in (\bigcup_{k=0}^r V^{(k)}) \setminus \{i\}} \pi_{\hat{\alpha}^{(t)}}(z_i, z_j)^{\ry_{i,j}}(1 - \pi_{\hat{\alpha}^{(t)}}(z_i,z_j))^{1 - \ry_{i,j}} \\
    &\cdot \mathbb{E}_{Z_i}\!\Big[\prod_{j \in \bigcup_{s=0}^{r-1} V^{(s)}}(1 - \pi_{\hat{\alpha}^{(t)}}(Z_i, z_j))\Big]^{\mathds{1}(i \notin V^{(r)})|\mathcal{U}|}, \numberthis \label{eq:full_cond}
\end{align*}
where $z_{-i}$ denotes the current latent positions of all sampled vertices other than $i$.

Both the stochastic E- and M-steps involve repeated evaluation of the intractable integral~\eqref{eq:lsm_exclusion_prob} for different values of $\theta$ and all $z_j \in V^{(0)} \cup \dots \cup V^{(r-1)}$.
The low dimensionality of the latent space ($d = 2$ or $3$ in practice) and the smoothness of the integrand enable efficient approximation via quasi-Monte Carlo (QMC) integration \citep{lemieux2008quasi}.
Specifically, we generate $2^M$ low-discrepancy points $u_1, \dots, u_{2^M}$ in $[0,1]^d$ using a Sobol sequence and map them to standard normal draws via the inverse transform.
At each sEM iteration, these fixed draws are scaled by $\sqrt{\psi}$ to produce samples $t_1, \dots, t_{2^M} \sim \mathcal{N}_d(0, \psi I_d)$.
Thus, rather than maintaining and updating $|\mathcal{U}|$ individual position vectors, the entire contribution of the unsampled vertices is approximated using only $2^M$ fixed quadrature points shared across all unsampled vertices, which in large networks with high $|\mathcal{U}|$ can yield a substantial computational saving.
The base low-discrepancy sequence is generated only once, and during the Gibbs update for $Z_i$, only the distances between $Z_i$ and the auxiliary points $t_1, \dots, t_{2^M}$ require re-computation, as distances to other sampled positions remain fixed.

The M-step is performed with the L-BFGS algorithm \citep{liu1989limited} using the derivatives with respect to $\alpha$ and $\log \psi$, the latter ensuring strict positivity of the variance parameter.

The sEM algorithm produces a sequence $\{\hat{\theta}^{(t)}\}_{t=1}^T$ whose stationary distribution is concentrated in a neighborhood of the maximum likelihood estimate of $\theta$ \citep{nielsen2000stochastic}.
After a sufficient burn-in period, the final parameter estimates are obtained by averaging over the tail of the sequence.
Detailed implementation of the Gibbs sampler with the ESS kernel, QMC integration, and the full sEM routine is provided in Appendix~\ref{A:5}.

\subsection{Simulations}
\label{sec:lsm_simulations}

A simulation study is conducted to evaluate the performance of the proposed sEM-based estimator of the distance model parameters under snowball sampling, and to compare it to the naive estimator, which is constructed by replacing \eqref{eq:complete_data_like} in the sEM algorithm with the complete data likelihood of the population model for the sampled vertices only, treating the snowball sample as a complete network:
\begin{equation} \label{eq:lsm_naive_like}
    f_\theta^{naive}(\ry, z_{V\setminus\mathcal{U}} \mid V^{(0)}) = \prod_{\{i,j\} \in (V\setminus\mathcal{U})^2} \pi_\alpha(z_i, z_j)^{\ry_{i,j}} (1 - \pi_\alpha(z_i,z_j))^{1 - \ry_{i,j}} \prod_{i \in V \setminus \mathcal{U}} f_\psi(z_i).
\end{equation}
In the rest of the paper, we refer to the snowball likelihood \eqref{eq:complete_data_like} as the `snowball-corrected model', and the naive likelihood \eqref{eq:lsm_naive_like} as the `naive model'.
Analogously, the resulting sEM-based parameter estimates are referred to as the `snowball-corrected' and the `naive' estimates, respectively.

\begin{figure}[htbp]
    \centering
    \begin{subfigure}{\linewidth}
        \centering
        \includegraphics[width = \linewidth]{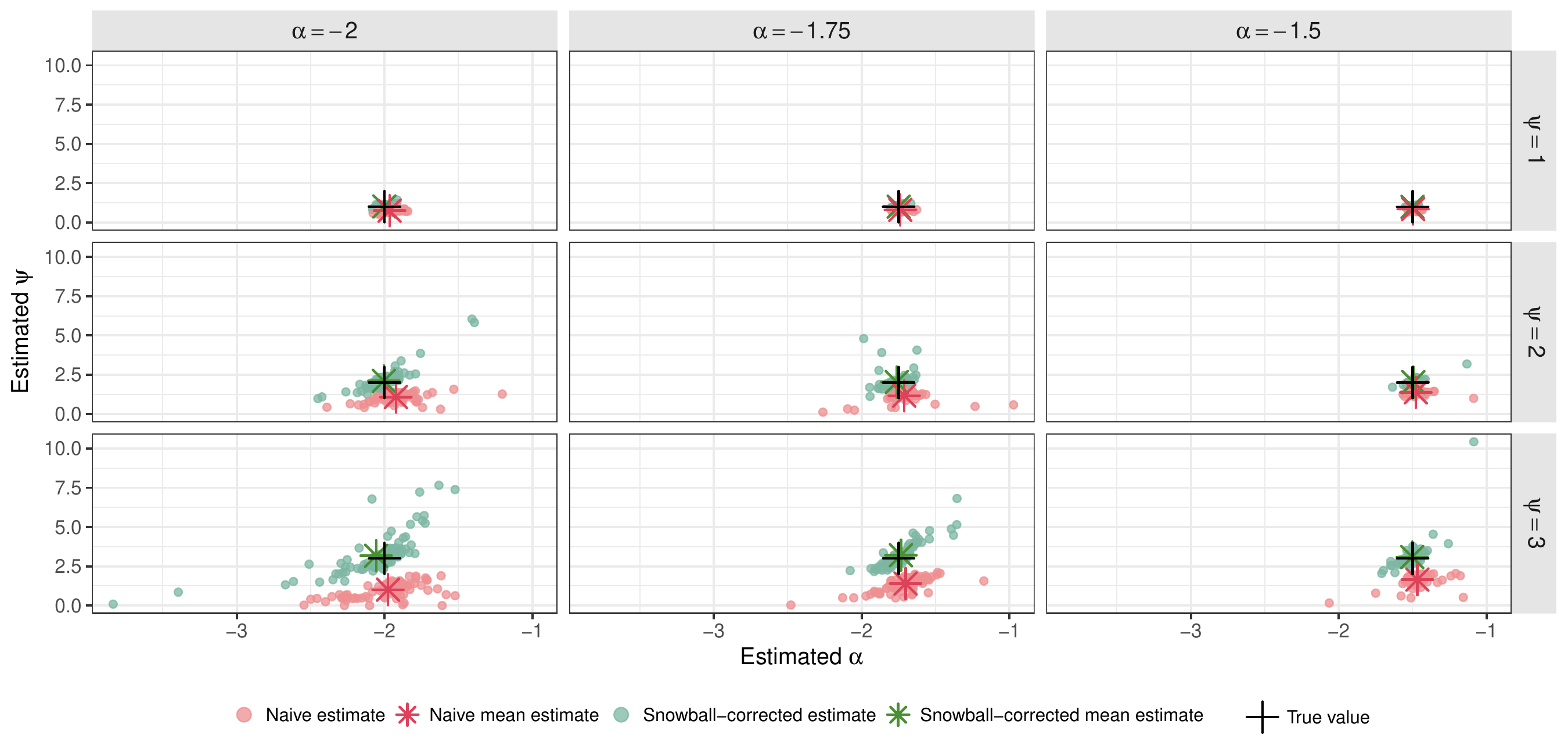}
        \caption{Distributions of $\alpha$ and $\psi$ estimates under the sEM estimator and the naive distance model.}
        \label{fig:lsm_sim_restricted_bayes}
    \end{subfigure}
    \begin{subfigure}{\linewidth}
        \centering
        \includegraphics[scale = 0.4]{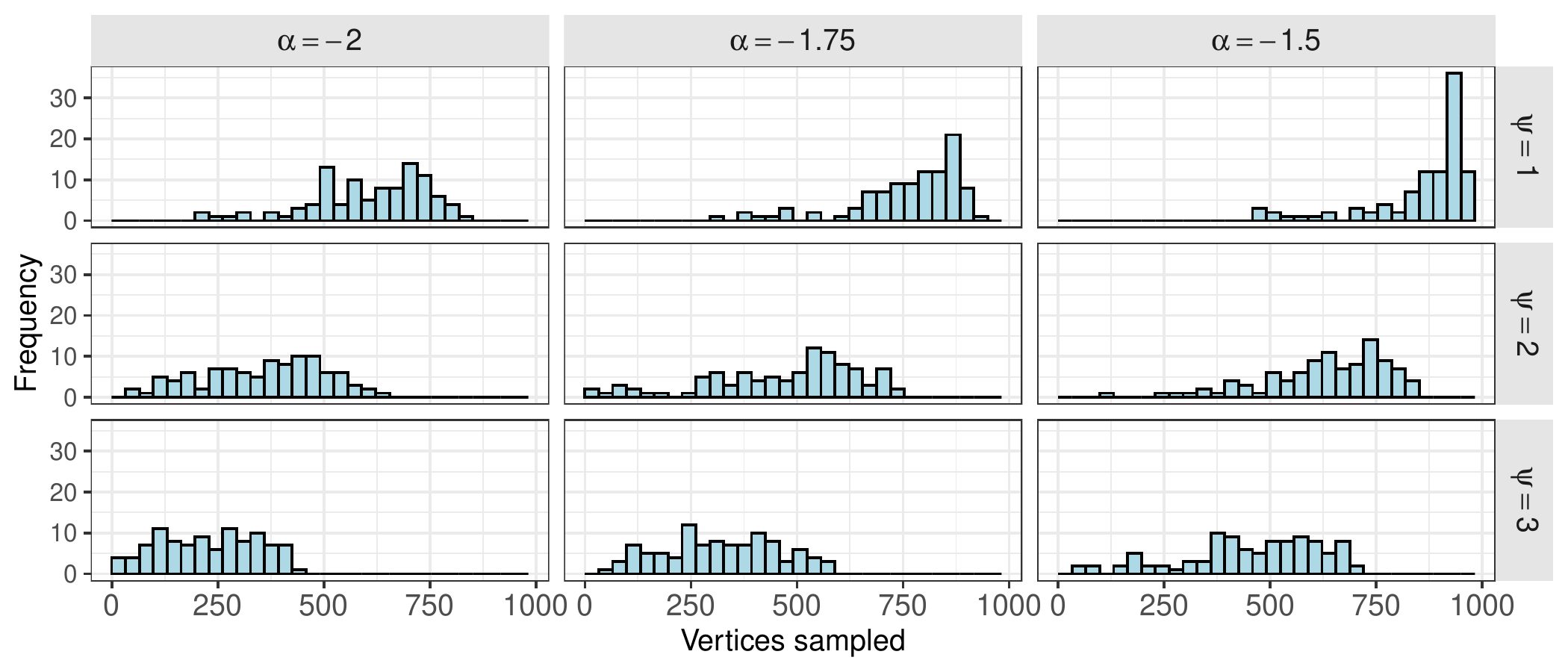}
        \caption{Distributions of the number of sampled vertices.}
        \label{fig:lsm_sim_sample_sizes}
    \end{subfigure}
    \caption{Simulation results for the distance model under 2-wave snowball sampling.
    Each panel corresponds to one of the $9$ parameter combinations $(\alpha, \psi)$, based on $100$ independent replications with population networks of size $N = 1{,}000$ and a two-dimensional latent space.}
\end{figure}

We generate synthetic networks of size $N = 1{,}000$ from the population model~\eqref{eq:hoff_model} with parameters $\alpha \in \{-2, -1.75, -1.5\}$ and $\psi \in \{1, 2, 3\}$ in a two-dimensional latent space ($d = 2$).
For each parameter combination $(\alpha, \psi)$, we simulate $100$ independent networks and draw a 2-wave snowball sample from a uniformly selected ego vertex in each network.
The parameter values are chosen based on preliminary experiments indicating that they yield snowball samples  covering a broad spectrum of sampling effort scenarios, as shown in Figure~\ref{fig:lsm_sim_sample_sizes}.
In rare cases where the snowball sample contains fewer than $3$ vertices or no triangles, we discard the sample and redraw, as such samples provide insufficient information for inference under the distance model.
The rate of such cases is about $0.2\%$ over all simulations.

For both the snowball-corrected and naive models, the sEM algorithm is run for a minimum of $30$ iterations, after which it continues for up to $1{,}000$ iterations or until the long-run averages of $\alpha$ and $\psi$ converge over $30$ consecutive iterations, whichever occurs first.
Final parameter estimates are obtained by averaging over the second half of the sEM sequence, excluding the initial $30$ iterations.
Depending on the size of the snowball sample, the number of MCMC iterations per sEM iteration is set adaptively between $100$ and $1{,}000$ via a heuristc rule (see Appendix~\ref{A:5} for details) to ensure sufficient mixing, while remaining computationally feasible across all simulations.
The larger snowball samples require fewer MCMC iterations due to faster convergence from the greater information content, while smaller samples result in a less concentrated posterior distribution of the latent positions and thus require more MCMC iterations to achieve sufficient mixing.
The sEM algorithm is initialized at $\hat{\alpha}^{(0)} = 0$ and $\hat{\psi}^{(0)} = 1$ for both models in all simulations, while the initial latent positions are set at the embeddings obtained from multidimensional scaling of the observed geodesic distances among the sampled vertices, as in \citet{hoff2002latent}.

Figures~\ref{fig:lsm_sim_restricted_bayes} and~\ref{fig:lsm_sim_sample_sizes} display the distributions of the parameter estimates and sample sizes across the $100$ simulations for each parameter combination.
The snowball-corrected estimates are well-centred on the true parameter values across all settings.
Larger values of $\psi$ generate more diffuse latent positions, producing sparser networks and smaller snowball samples, as shown in Figure~\ref{fig:lsm_sim_sample_sizes}.
In such cases, the variance of the snowball-corrected estimates increases, though the bias remains near zero for all parameter combinations except $(\alpha, \psi) = (-1.5, 3)$, where a slight negative bias in $\hat{\alpha}$ is observed.
This is due to the possibility of extremely small snowball samples under such parameter combination, which drives $\hat{\psi}$ close to zero, while pulling $\hat{\alpha}$ downward to compensate.
Even then, the bias of the snowball-corrected estimates remains substantially smaller than that of the naive estimates, which underestimate $\psi$ across most parameter settings.
The naive estimates of $\alpha$ largely agree with those of the snowball-corrected model with a slight exception at $\alpha = -2$ and $\phi > 1$, where the naive estimator exhibits a small positive bias.
Only at $\psi = 1$, where snowball samples cover more than $70\%$ of the population on average, does the naive estimator match the performance of the snowball-corrected estimator, as the large sampling fraction renders the bias from ignoring the sampling design negligible.

\section{Application: Semiconductor Patent Co-inventorship Network}
\label{sec:application}

We apply the snowball-corrected distance model to a co-inventorship network constructed from patent filings in the field of semiconductors submitted to the European Patent Office or the German Patent and Trademark Office (\textit{Deutsches Patent- und Markenamt}) between 1997 and 2012.
The underlying patent-inventor dataset is described in detail in \citet{fritz2023modelling}, and we restrict attention to the semiconductors sub-area to obtain a well-defined co-inventorship network of manageable scale.
Two inventors are linked by an undirected edge if they appear as co-inventors on at least one patent within the study period, with multiple co-invented patents contributing at most one edge.
After retaining only the largest connected component, the resulting network has $N = 5{,}979$ vertices and $21{,}073$ edges with mean degree $7.09$, maximum degree $96$, average shortest path length $7.55$ and a global clustering coefficient of $0.31$.

Fitting the distance model directly to the full network is computationally intractable at $N = 5{,}979$, and the snowball-corrected model developed in this paper removes that barrier.
This application is also a demanding one, as the network exhibits `small-world' characteristics, i.e. high clustering and low average shortest path length \citet{watts1998collective}, as well as a heavy-tailed degree distribution, typical of scientific collaboration networks \citep{newman2001structure}.
Appendix~\ref{A:6} reports a small-world measure of $0.11$ as defined by \citet{telesford2011ubiquity}, indicating the small-world nature of the network, along with a skewed degree distribution with a long right tail driven by a small number of highly connected inventors who co-author with many others and a large number of peripheral inventors with few co-inventions.
\citet{rastelli2016properties} show theoretically that the latent position model with a Gaussian connection function, which is closely related to the distance model of \citet{hoff2002latent}, produces a vanishing clustering coefficient and thin-tailed degree distribution as the network grows, and therefore cannot represent the small-world behaviour or heavy-tailed degree distributions commonly observed in social networks.
Our application to the semiconductor co-inventor network therefore constitutes the first empirical stress test of the distance model on a large-scale network exhibiting genuine small-world properties, which are only meaningfully present at a scale that has historically been computationally inaccessible to MCMC-based inference.
As shown in Section~\ref{sec:application_gof}, even in this unfavourable setting the snowball-corrected model recovers substantially more network structure than the naive approach.

Each inventor is associated with two covariates used in Section~\ref{sec:application_latent} to interpret the estimated latent positions: inferred gender and geographic distance. 
Each inventor has a recorded residence address from which longitude and latitude coordinates are derived \citep{fritz2023modelling}.
Pairwise geographic distances are then computed via the haversine formula.
Gender is not directly recorded in the patent register and is imputed from first names using a probabilistic name-gender matching procedure and is thus subject to some uncertainty.
The proportion of female inventors is below $5\%$ for the resulting network, consistent with the broader electrical engineering population studied by \citet{fritz2023modelling}.

\subsection{Sampling Design and Estimation}
\label{sec:application_sampling}

We draw $500$ independent snowball samples from the population network, each initiated from a single ego selected uniformly at random from $V$.
For each sample, the number of waves is chosen to be as large as possible subject to the constraint that the total number of sampled vertices does not exceed $1{,}000$ vertices to keep estimation computationally feasible for each sample.
The resulting sample sizes are highly heterogeneous, reflecting the pronounced degree heterogeneity of the network.
Samples initiated from high-degree hubs expand rapidly and saturate the vertex budget within one or two waves, producing dense, well-connected subgraphs, while samples initiated from peripheral vertices may require more waves to approach the budget and can yield sparse, tree-like subgraphs with limited inferential content.

For each of the $500$ samples, we fit the snowball-corrected model along with the naive model as a benchmark, obtaining from each model a pair of parameter estimates $(\hat{\alpha}^{(b)}, \hat{\psi}^{(b)})$ for $b = 1, \dots, 500$.
The sEM algorithm for the naive model is initialized at $\alpha^{(0)} = 0$ and $\psi^{(0)} = 1$, while for the snowball-corrected model, the sEM algorithm is initialized at the naive estimates for each sample to facilitate convergence.
Similar to the simulation study in Section~\ref{sec:lsm_simulations}, the initial latent positions are set via multidimensional scaling and the number of MCMC iterations per sEM iteration is set adaptively between $100$ and $1{,}000$ based on the size of the snowball sample.
To ensure reliable estimation for each sample, the sEM algorithm is run for $600$ iterations including a burn-in of $100$ iterations for both models.

\subsection{Aggregating Estimates Across Samples}
\label{sec:application_aggregate}
 
\begin{figure}[ht]
    \centering
    \includegraphics[scale = 0.525]{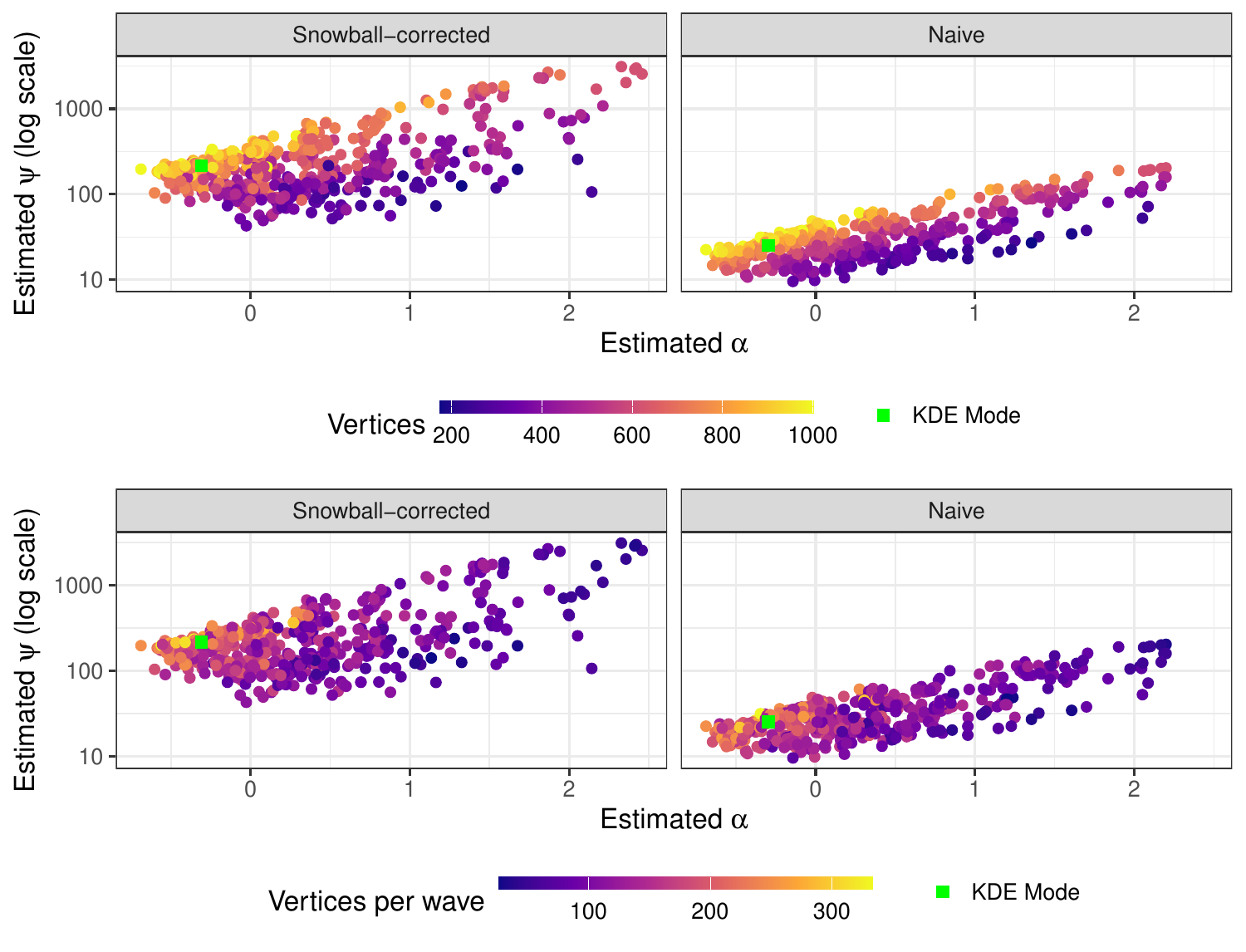}
    \caption{Pairs of estimates $(\hat{\alpha}^{(b)}, \hat{\psi}^{(b)})$ obtained by applying the snowball-corrected (left) and naive (right) models to each of the $500$ snowball samples, with $\hat{\psi}$ on a log scale. 
Top row: estimates colour-coded by total number of vertices sampled; bottom row: colour-coded by number of vertices sampled per wave. 
The green square marks the KDE mode used as the aggregate estimate.}
    \label{fig:real_estimates}
\end{figure}

The $500$ pairs of estimates $(\hat{\alpha}^{(b)}, \hat{\psi}^{(b)})$ from both the snowball-corrected and naive models are displayed in Figure~\ref{fig:real_estimates}, colour-coded by total vertices sampled and by vertices sampled per wave.

The snowball-corrected estimates form a heterogeneous cloud with a pronounced right tail in both parameters, driven by sparse, tree-like samples that carry limited information about the global connectivity structure of the network.
The naive estimates occupy a different region of the parameter space, with systematically lower $\hat{\psi}$ values, and exhibit less variability across samples.

A point estimate of the population parameters can be obtained by aggregating the estimates across the $500$ samples.
The aggregation of estimates across different samples from the same population is a common problem in meta-analysis, where the goal is to combine estimates from multiple independent studies to obtain an overall estimate of the population parameter.
The classical approach is the inverse-variance weighted mean, which is optimal under approximate normality of the individual estimates around the true parameter value \citep{hedges1992meta}.
This is equivalent to feasible generalized least squares regression of the estimates on a constant, with the inverse of the estimated covariance matrix of the estimates as weights, as performed by \citet{stivala2016snowball} in their aggregation of estimates in their application of the conditional estimation of ERGMs to snowball samples.

In our setting, this approach is inadvisable on two grounds.
First, the heterogeneous covariances of the sampling distributions of both snowball-corrected and naive estimates are of unknown form.
Second, the sampling distribution is highly non-normal, with a pronounced right skew and heavy tails, particularly for $\hat{\psi}$.
These difficulties are analogous to those arising in meta-analysis when study-level estimates are heterogeneous with the presence of outliers.
\citet{hartwig2020median} proposes a mode-based estimator under the zero modal bias assumption (ZEMBA) as a robust meta-analytic estimator in such settings.
ZEMBA states that the most common value of the bias across samples is zero, so that the mode of the distribution of estimates is consistent for the true parameter even when the majority of individual estimates are biased.
The plausibility of ZEMBA in our setting is directly supported by the simulation results of Section~\ref{sec:lsm_simulations}.
The mode of the snowball corrected estimates across samples closely tracks the true parameter values, confirming that sufficiently well-connected samples, which constitute the majority of the $500$ draws, yield approximately unbiased estimates.

Similar to the approach of \citet{hartwig2020median}, we aggregate the estimates across samples by fitting a bivariate kernel density estimate (KDE) to the joint distribution of $(\hat{\alpha}^{(b)}, \hat{\psi}^{(b)})$ with log-transformation of $\hat{\psi}^{(b)}$ to mitigate the right skew.
The bandwidth matrix is selected by the multivariate plug-in selector of \citet{wand1994multivariate}, as implemented in the \texttt{ks} package \citep{duong2007ks}, with pilot bandwidths estimated via the sum of asymptotic mean squared error method of \citet{duong2003plugin}.
The KDEs for both the snowball-corrected and naive estimates are weighted by the number of vertices sampled per wave for each replicate, upweighting samples that provide more information per unit of sampling effort and downweighting the sparse, chain-like samples responsible for the outlying estimates.
The aggregate estimate is the mode of the fitted KDE, with $\hat{\psi}^{(b)}$ back-transformed via exponentiation.
This procedure yields aggregate estimates of $(\hat{\alpha}, \hat{\psi}) = (-0.31,\,215.17)$ for the snowball-corrected model and $(\hat{\alpha}_{\mathrm{naive}}, \hat{\psi}_{\mathrm{naive}}) = (-0.3, 25.17)$ for the naive model.
An $8.5$ times lower $\psi$ estimate shows that the naive model estimates a much more compressed latent arrangement of the vertices, which directly reflects the upward bias in edge probability induced by ignoring the snowball sampling design, similar to the bias observed in the simulation study of Section~\ref{sec:lsm_simulations} and, for the Erd\H{o}s--R\'{e}nyi case, by \citet{sapargali2026exact}.
The close agreement in $\hat{\alpha}$ between the two models is also consistent with the simulation results, where $\hat{\alpha}$ are largely the same across the two models.

\subsection{Goodness of Fit}
\label{sec:application_gof}

To assess model fit, we draw networks from the population posterior predictive distribution under each of the two aggregate parameter estimates and compare them to the observed network using a suite of diagnostic statistics.
More precisely, using the aggregate parameter estimates from the snowball-corrected and naive models, we draw for each model $7{,}000$ matrices of latent positions from the population posterior $Z_1, \dots, Z_{5979} \mid Y = y, \theta$ via the ESS-within-Gibbs sampler.
By discarding the first $2{,}000$ iterations as burn-in and thinning by a factor of $10$, we obtain $500$ approximately independent draws from the posterior distribution of latent positions, which are then used to generate $500$ networks from the population model.
From these networks, we compute the distribution of each diagnostic statistic to assess how well the snowball-corrected model captures different aspects of the observed network structure.

Following \citet{hunter2008gof}, traditionally such assessment is performed by overlaying the distribution of the observed statistic on boxplots of the same statistic computed across networks simulated from the fitted model.
For a network with $5{,}979$ vertices, such boxplots become visually cluttered and difficult to read, as the support and variability of the diagnostic statistics becomes large.
We instead display ribbon plots showing the mean simulated empirical cumulative distribution function (ECDF) together with a shaded band spanning the minimum to maximum across the $500$ simulations, with the observed ECDF overlaid.
We consider three diagnostic statistics chosen to capture network structure at increasing levels of vertex aggregation: the degree distribution (local), the edgewise shared partner (ESP) distribution (mesoscale), and the Laplacian spectrum (global), assessed via the spectral goodness-of-fit statistic (SGOF) of \citet{shore2015spectral}.

As a preliminary check of the basic fit, Table~\ref{tab:gof_basic} reports the mean number of edges and mean number of isolates across the $500$ simulations, alongside the observed values.
The observed network has $21{,}073$ edges and, by construction, zero isolates since we retain only the largest connected component.
The snowball-corrected model reproduces the observed edge count closely, with a mean of $24{,}092.79$ edges.
The naive model fails this basic test, producing networks with a mean of $41{,}154.16$ edges, roughly $1.71$ times the observed count, which is a direct consequence of the compressed latent space implied by the naive estimates yielding inflated edge probabilities across all pairs of vertices.
Both models produce networks with a non-zero mean number of isolates ($137.98$ and $24.35$ respectively), reflecting a known limitation of the distance model: the logistic connection probability decays to zero at large latent distances, so peripheral vertices in the latent configuration are rarely connected to any other vertex.
In this context, the higher mean number of isolates under the snowball-corrected model is a result of its lower edge density.
This has direct implications for the ESP comparison below.

\begin{table}[t]
    \centering
    \caption{Mean number of edges and isolates in $500$ networks simulated from the posterior predictive distribution under the snowball-corrected and naive models. The observed network has $21{,}073$ edges and zero isolates.}
    \label{tab:gof_basic}
    \begin{tabular}{lcc}
        \toprule
        & Mean edges & Mean isolates \\
        \midrule
        Snowball-corrected & $24{,}092.79$ & $137.98$ \\
        Naive & $41{,}154.16$ & $24.35$ \\
        \bottomrule
    \end{tabular}
\end{table}

\begin{figure}[ht]
  \centering
  \includegraphics[width = \textwidth]{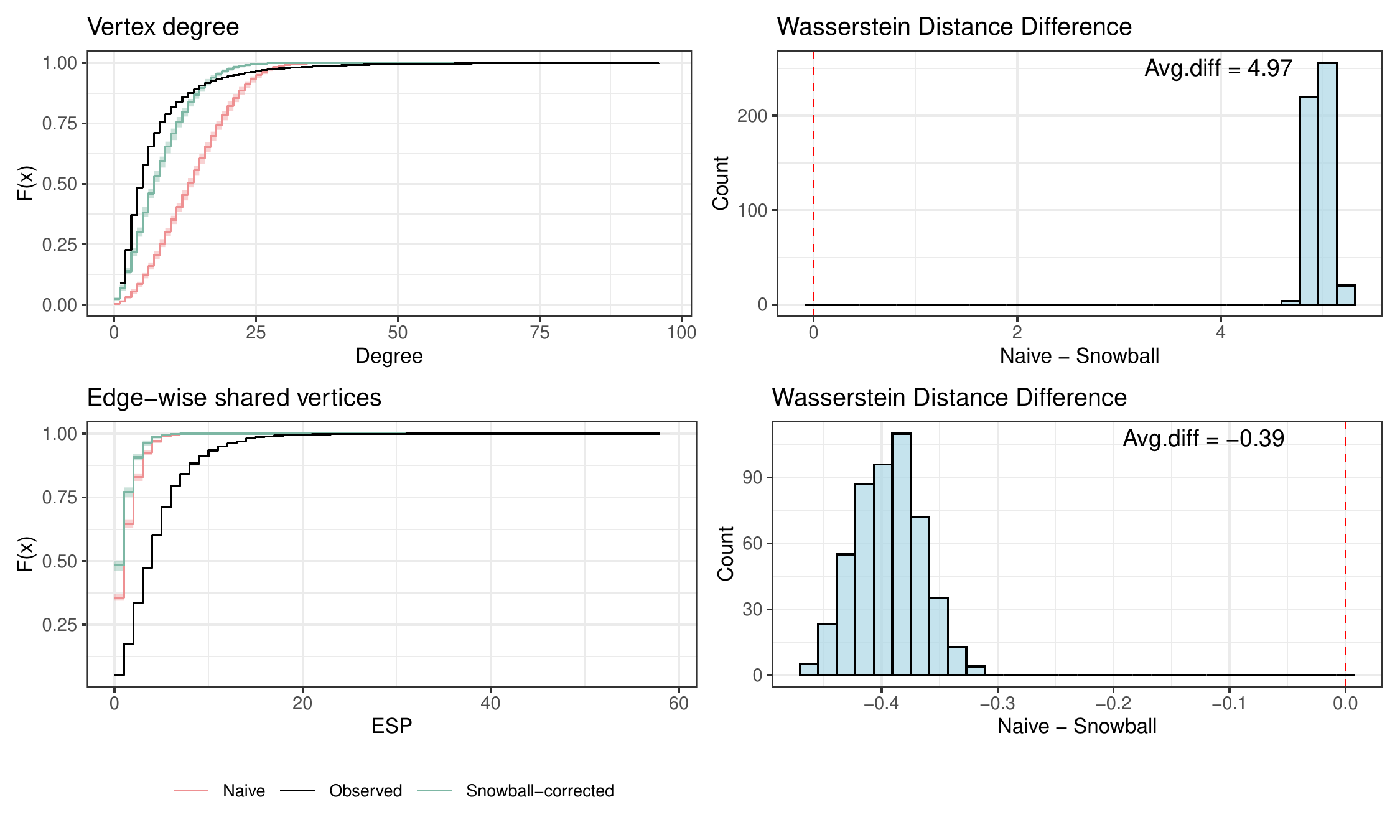}
  \caption{Goodness-of-fit diagnostics for the degree distribution (top row) and edgewise shared partner (ESP) distribution (bottom row).
  Left panels: ribbon plots of the simulated ECDF under the snowball-corrected and naive models across $500$ posterior predictive draws, with the observed ECDF overlaid (black).
  The shaded ribbon spans the minimum to maximum simulated ECDF across the $500$ draws.
  Right panels: histograms of the per-simulation difference in Wasserstein distance to the observed ECDF between the distributions under naive and snowball-corrected models; positive values favour the snowball-corrected model and the red dashed line marks zero.}
  \label{fig:gof_ecdf}
\end{figure}

\paragraph{Degree distribution.}
The top row of Figure~\ref{fig:gof_ecdf} displays the ECDF of the degree distribution for the observed network alongside the ribbon plots of the simulated ECDFs under each model, together with histograms of the difference in Wasserstein distance to the observed ECDF between the simulated distributions under the naive and snowball-corrected models.
The snowball-corrected model tracks the observed degree ECDF closely throughout its range.
The histogram in Figure~\ref{fig:gof_ecdf} (top right) lies entirely to the right of zero for all simulations, with an average difference of $4.97$ in favour of the snowball-corrected estimator.
The naive model systematically underestimates the proportion of low-degree vertices and overestimates the proportion of high-degree vertices, which is again a consequence of the inflated edge density producing networks that are globally too well-connected.

\paragraph{Edgewise shared partners.}
The bottom row of Figure~\ref{fig:gof_ecdf} displays the analogous diagnostics for the ESP distribution.
The Wasserstein distance difference reverses here as the naive model achieves a smaller distance to the observed ESP distribution by an average of $0.39$ across simulations, with the histogram lying entirely to the left of zero.
This apparent advantage of the naive model however is simply an artifact of the inflated edge density documented in Table~\ref{tab:gof_basic}.
A denser network mechanically generates more shared partners per edge.
The naive model's excess edges push its ESP distribution closer to the observed one in Wasserstein distance, despite the underlying generative mechanism being misspecified.
Given the known limitations of the distance model with small-world networks and a more accurate edge count, the edge-wise shared partner distribution of the snowball-corrected model provides a more faithful account of the clustering structure fitted by the distance model.

\begin{figure}[ht]
  \centering
  \includegraphics[width = 0.625\textwidth]{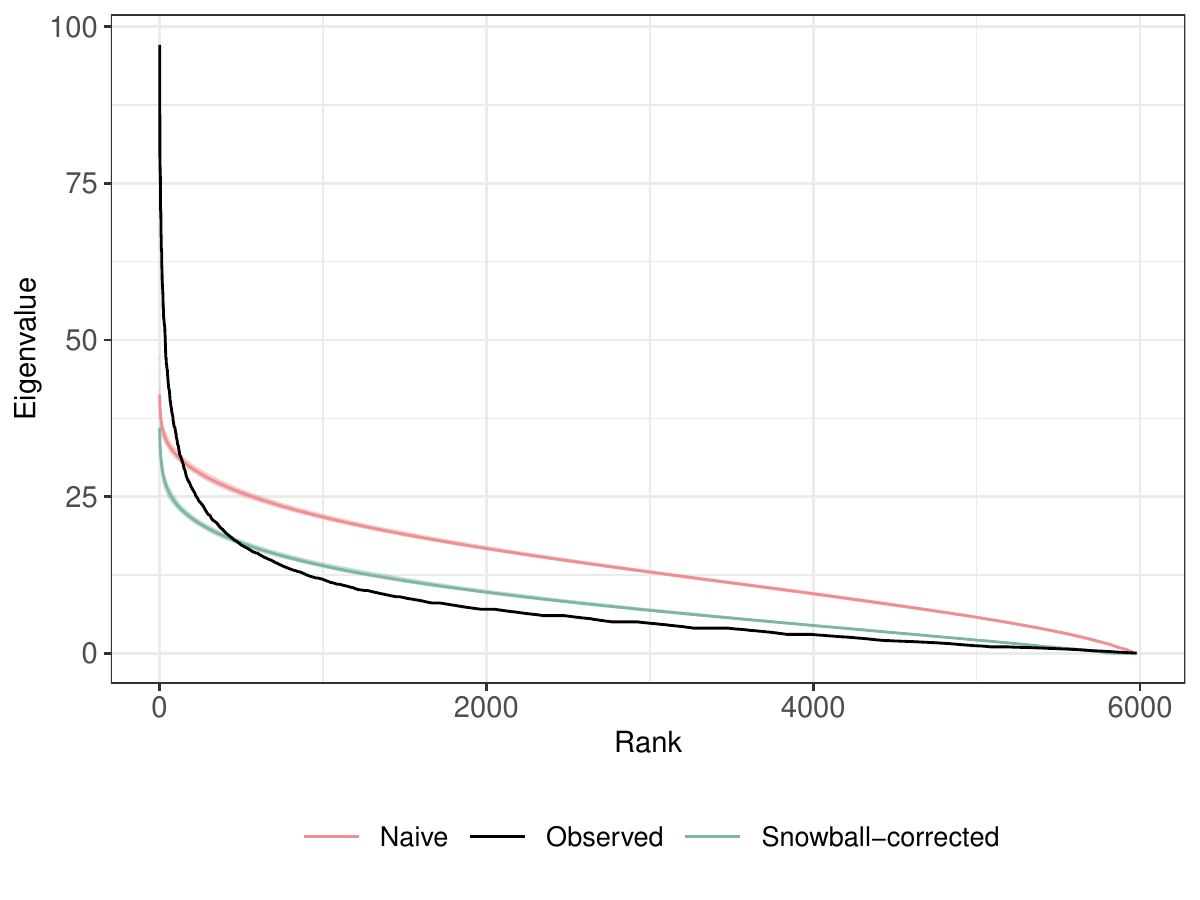}
  \caption{Ribbon plots of the ranks of eigenvalues of the simulated Laplacians under the snowball-corrected and naive models for $500$ posterior predictive draws with the observed ranked eigenvalues overlaid.
  The shaded ribbon spans the minimum to maximum simulated eigenvalues across the $500$ draws.}
  \label{fig:gof_eigen}
\end{figure}

\paragraph{Laplacian spectrum.}
Figure~\ref{fig:gof_eigen} displays the ribbon plots of the ranks of eigenvalues of the simulated Laplacians under each model, with the observed ranked eigenvalues overlaid.
The snowball-corrected model tracks the bulk of the observed spectrum much more closely than the naive model, which overshoots the observed eigenvalues throughout the mid-range.
The only exception are the few leading eigenvalues.
The naive model achieves top eigenvalues closer to the observed value than the snowball-corrected model.
This is again due to the higher edge density, as the leading eigenvalue of the graph Laplacian scales approximately with the maximum degree \citep{chung1997spectral}, and the naive model's denser networks produce higher maximum degrees, inflating the eigenvalues without any genuine improvement in global structural fit.

Table~\ref{tab:sgof} quantifies global spectral fit via the SGOF statistic of \citet{shore2015spectral}, which measures the relative improvement in Laplacian spectral fit over an Erd\H{o}s--R\'{e}nyi null model, with zero indicating no improvement and one indicating perfect fit.
The naive model achieves an SGOF of $0.027$ (90\% interval: $0.019$--$0.036$) demonstrating tiny improvement over the ER null model.
In contrast, the snowball-corrected model exhibits an SGOF of $0.258$ (90\% interval: $0.241$--$0.274$) yielding a ninefold improvement over the naive model.
The two intervals do not overlap, confirming that the improvement from correcting for the sampling design is unambiguous.
Nevertheless, the modest SGOF of $0.258$ for the snowball-corrected model reflects the known difficulty of fitting Euclidean latent space models to small-world networks with heavy-tailed degree distributions, as discussed above.

\begin{table}[t]
\centering
\caption{Spectral goodness-of-fit (SGOF) of \citet{shore2015spectral} for networks simulated from the posterior predictive distribution under the snowball-corrected and naive models.
The 90\% interval is computed across $500$ simulations.}
\label{tab:sgof}
\begin{tabular}{lccc}
\toprule
 & SGOF & 5th percentile & 95th percentile \\
\midrule
Snowball-corrected & $0.258$ & $0.241$ & $0.274$ \\
Naive & $0.027$ & $0.019$ & $0.036$ \\
\bottomrule
\end{tabular}
\end{table}

\subsection{Latent Position Analysis}
\label{sec:application_latent}

\begin{figure}[htbp]
    \centering
    \begin{subfigure}{\linewidth}
        \centering
        \includegraphics[width = \linewidth]{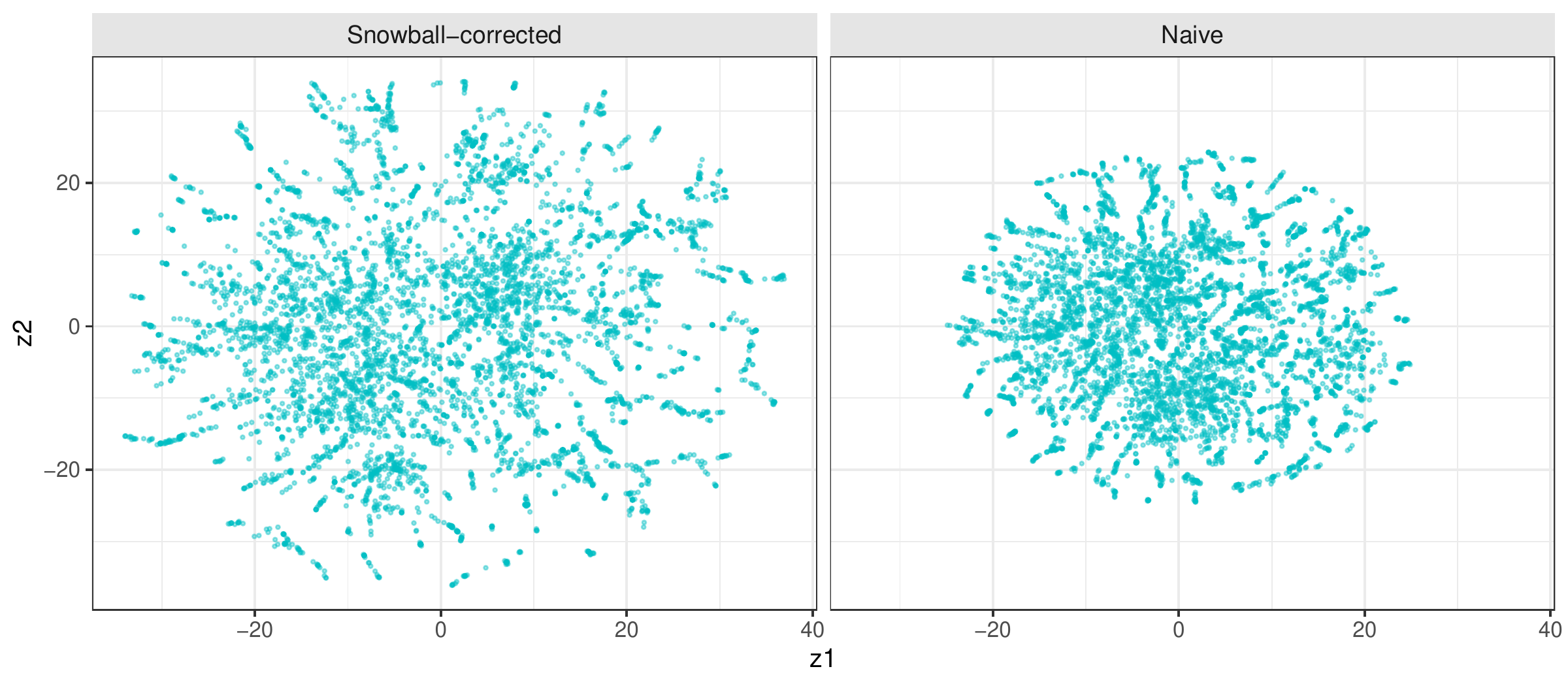}
        \caption{Posterior means of Procrustes-aligned latent positions under the snowball-corrected (left) and naive (right) models by inventor's gender, across $500$ approximately independent draws from the population posterior $Z_1, \dots, Z_{5979} \mid Y = y, \theta$, aligned to the draw with the highest complete-data log-likelihood.}
        \label{fig:lat_pos}
    \end{subfigure}
    \begin{subfigure}{\linewidth}
        \centering
        \includegraphics[width = \linewidth]{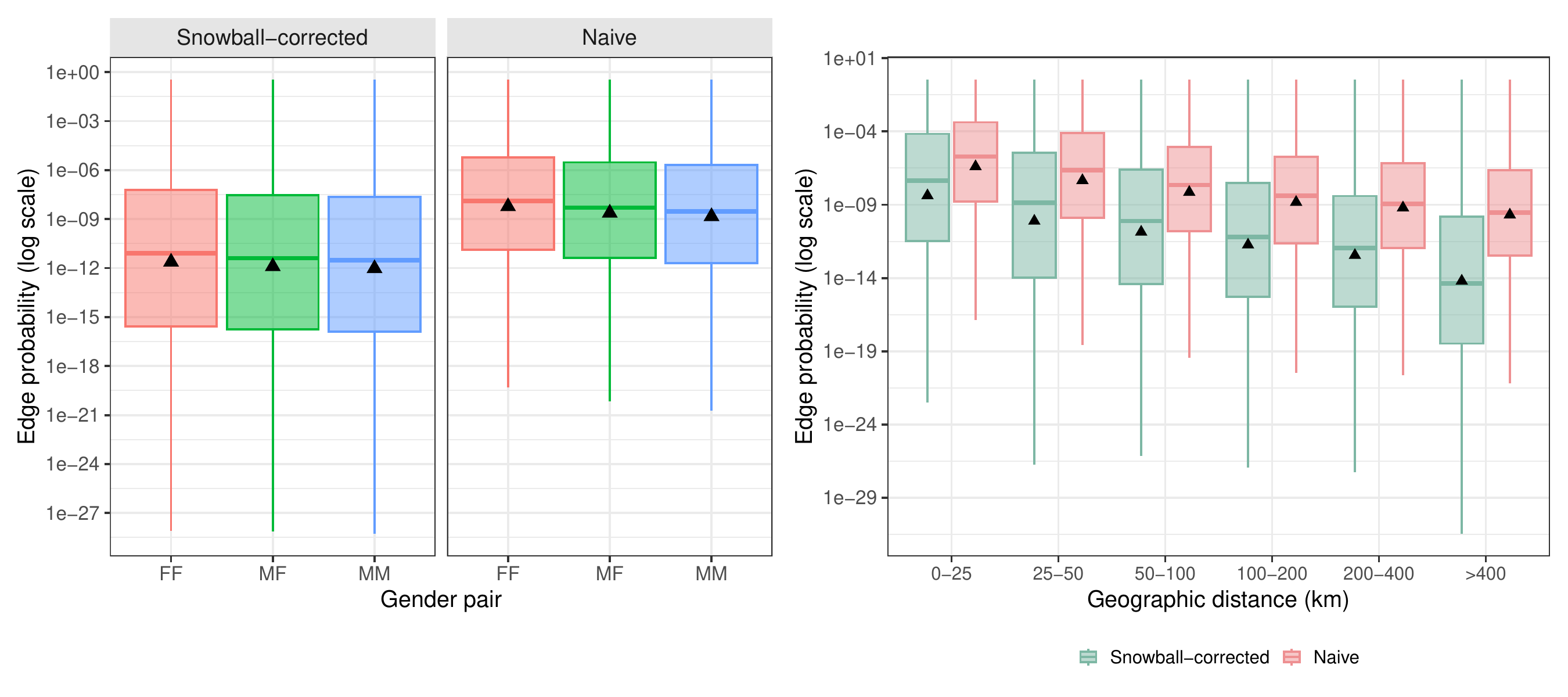}
        \caption{Implied edge probabilities under the snowball-corrected and naive models across $500$ posterior draws.
        Left: distribution of posterior edge probabilities across all dyads within each gender pair (FF, MF, MM), with the triangle marking the mean.
        Right: distribution of posterior edge probabilities across all dyads within each geographic distance bin, with the triangle marking the mean.
        Both axes use a log scale.}
        \label{fig:cov_effects}
    \end{subfigure}
    \caption{Latent position estimates and implied covariate effects under the snowball-corrected and naive models.
    Mean edge probabilities are higher under the naive model across all groups, reflecting the inflated edge density of its compressed latent configuration.}
    \label{fig:latent_analysis}
\end{figure}

Figure~\ref{fig:lat_pos} displays the posterior means of the Procrustes-aligned latent positions under each model, serving as the geometric summary of the inferential difference between the two approaches.
The snowball-corrected model distributes inventors across a substantially wider region of the latent space than the naive model, whose positions are compressed into a markedly tighter configuration around the origin.
With $\hat{\psi}_{\mathrm{naive}} = 25.17$ compared to $\hat{\psi} = 215.17$ under the corrected model, the naive model pulls all positions toward a denser, more centralised configuration.
The consequence, as documented in Table~\ref{tab:gof_basic}, is a posterior predictive edge count nearly twice the observed value.
The compression of the latent space under the naive model has direct implications for the implied edge probabilities across all dyads, along gender homophily and geographic proximity, as we demonstrate below.

\paragraph{Gender.}
The left panel of Figure~\ref{fig:cov_effects} displays the distribution of posterior edge probabilities across all dyads within each gender pairing.
The mean edge probabilities for female--female (FF), mixed-female (MF), and male--male (MM) pairs decrease monotonically in that order under both models, corroborating the gender homophily finding of \citet{fritz2023modelling}, who identify female inventors as significantly more likely to collaborate with other female inventors than with males, while male inventors show no analogous preference.
The mean FF to MM ratio of edge probabilities is $1.40$ under the snowball-corrected model and $1.42$ under the naive model, indicating that the two models agree closely on the direction and relative magnitude of gender homophily.
However, the absolute edge probabilities differ substantially between models.
Mean FF pair probabilities are $1.09 \times 10^{-3}$ under the snowball-corrected model compared with $1.77 \times 10^{-3}$ under the naive model, an inflation of approximately $60\%$ that applies uniformly across all gender pairings.
Furthermore, the naive model's compressed latent space underestimates the variability in edge probabilities within each gender pairing, as reflected in the narrower distributions in the left panel of Figure~\ref{fig:cov_effects}.
A researcher relying on the naive model would thus arrive at the same qualitative conclusions about gender homophily, but would systematically overestimate the baseline rate of inventor collaboration and underestimate the heterogeneity in collaboration propensities across inventor pairs.

\paragraph{Geographic proximity.}
The right panel of Figure~\ref{fig:cov_effects} shows a pronounced monotone decay in posterior edge probabilities with geographic distance under both models, replicating the positive and significant spatial proximity effect of \citet{fritz2023modelling}.
Across six distance bins ranging from below $25$\,km to above $400$\,km, edge probabilities decline by roughly four orders of magnitude.
Similar to the gender pairing, the naive model underestimates the variability in edge probabilities within each distance bin.
More importantly, the decay in mean edge probabilities is consistently steeper under the snowball-corrected model.
The ratio of mean edge probabilities between the closest bin ($0$--$25$\,km) and the most distant bin ($>400$\,km) is $22.3$ under the corrected model compared with $14.4$ under the naive model.
This amplification is consistent with the snowball-corrected model's more dispersed latent configuration, in which geographic separation between inventors translates into a larger latent distance penalty.

The implied edge probabilities under the snowball-corrected model indicate that geographically proximate pairs are substantially more likely to collaborate, and female--female pairs exhibit the strongest gender homophily on average.
While the naive model recovers the same qualitative orderings, its compressed latent space systematically overstates edge probabilities across all groups and attenuates the geographic distance effect, with the proximate-to-distant ratio falling from $22.3$ to $14.4$.
Correcting for the snowball sampling design thus matters not only for reproducing the global structure of the network, as demonstrated in Section~\ref{sec:application_gof}, but also for the accurate quantification of the dyad-level mechanisms driving inventor collaboration.

\section{Discussion}
\label{sec:discussion}

This paper derives the exact likelihood of a multi-wave snowball sample for the class of CLS models and demonstrates that the conditional independence of edge formation, given latent vertex-level quantities, reduces an otherwise intractable marginalization over unobserved network configurations to a closed-form expression in the model parameters and the observed wave sets.
Starting from the ER model, we extended the framework to the full CLS class and provided a concrete implementation for the distance model of \citet{hoff2002latent} via a stochastic EM algorithm.
Simulation studies and the application to the semiconductor co-inventor network confirmed, at a scale previously inaccessible to likelihood-based latent space inference, that ignoring the snowball sampling mechanism produces misleading estimates.
The naive procedure severely underestimates the spread of the latent space, inflates edge probabilities, and, in the case of the semiconductor network, achieves a spectral goodness-of-fit nine times worse than the corrected model.

The most significant theoretical limitation of the present framework concerns the scope of the CLS class itself.
The closed-form reduction at the heart of our result depends critically on edges forming independently across vertex pairs conditional on the latent positions.
This assumption fails for models in which edge probabilities depend on the joint configuration of multiple dyads, most prominently ERGMs, where the sufficient statistics couple dyads in a way that prevents the marginalization over unobserved configurations from factorizing.
For such models the conditional maximum likelihood approach of \citet{pattison2013conditional} remains a viable route to snowball-corrected inference, and extending the exact likelihood correction to dyad-dependent models remains an open problem.

A second limitation concerns non-ignorable ego selection, treated in Section~\ref{subsec:cls_extension}.
When ego vertices are selected based on network features such as degree, centrality, or community membership, which depend on the latent positions, the selection mechanism is non-ignorable and must be modelled explicitly, at considerable computational cost even in the simplest cases.
As discussed in Section~\ref{subsec:cls_extension}, this difficulty is not unique to CLS models, and no existing model-based approach to snowball sampling provides a general treatment of it.
A general solution within the CLS framework would require deriving the adjusted prior $f(z \mid V^{(0)})$ for each application-specific selection mechanism, while a unified treatment remains an open problem.

The application to the semiconductor co-inventor network also empirically reveals a structural limitation of the Euclidean latent distance model that is independent of the sampling correction.
Even under the corrected likelihood, the small-world characteristics and heavy-tailed degree distribution of the network pose a challenge, as theoretically predicted by \citet{rastelli2016properties}.
The corrected model fits substantially better than the naive one, but the absolute goodness-of-fit remains modest.
Importantly, this incompatibility is more honestly exposed by the corrected procedure than by the naive one.
The naive estimator artificially contracts the latent space, distorting the latent geometry, which results in a substantially worse fit.
This observation motivates the extension of the snowball-corrected likelihood to richer model specifications within the CLS class that are better suited to networks with small-world characteristics and heavy-tailed degree distributions.
Hyperbolic latent space models \citep{krioukov2010hyperbolic}, which embed vertices in a negatively curved space that naturally accommodates heavy-tailed degree distributions, and graphon models, which allow degree heterogeneity to be captured without imposing a parametric form on the latent position distribution, are particularly natural candidates.
For distance models, one can add vertex-specific random effects to the linear predictor to capture degree heterogeneity \citep{kaur2023latent} and small-world structure.
All of these satisfy the conditional edge independence requirement of the CLS class, meaning that the likelihood derivation of Section~\ref{subsec:cls_extension} applies without much modification.

The framework also extends to models with covariates.
If the covariates are observed for all vertices in the population, or for all dyads in the case of dyad-level covariates, they can be incorporated directly into the linear predictor of the CLS model without modification.
When covariates involving unsampled vertices are unobserved, a potential approach is to operationalize the missingness through a distributional assumption on the covariate values, provided the missingness mechanism is ignorable \citep{ibrahim1990incomplete}.
The expected exclusion probability integral then extends to a joint expectation over both latent positions and missing covariates.

Finally, a fully Bayesian implementation of the snowball-corrected likelihood is a natural extension of the present framework.
The sEM algorithm treats $\alpha$ and $\psi$ as fixed and estimates them by maximum likelihood.
A Bayesian implementation would propagate uncertainty in these parameters directly into the posterior over latent positions.
Crucially, this extension does not introduce substantial additional computational overhead.
The quasi-Monte Carlo approximation to the exclusion probability integral scales with the number of Sobol points and the dimension of the latent space, not with the number of unsampled vertices, so re-evaluating it at each posterior draw of $(\alpha, \psi)$ adds no more cost per iteration than the stochastic E- and M-steps of the sEM algorithm.
The Bayesian extension therefore remains substantially cheaper than full-population MCMC, which scales quadratically with $N$.

Taken together, these extensions suggest that the framework developed here can serve as a broadly applicable foundation for model-based inference on snowball-sampled network data.
The key architectural property is the separation between the sampling correction, which depends only on the wave structure and the conditional independence assumption, and the model specification, which can be varied freely within the CLS class.
Progress on either front, whether through better models for heterogeneous networks or more flexible treatments of non-ignorable selection, integrates immediately into the existing framework.
More broadly, the result points towards a general principle: when a model's likelihood factorizes over dyads, the snowball sampling mechanism can be accounted for at little additional cost, encouraging broader adoption of principled, sampling-aware inference in large network analysis.

\bibliographystyle{chicago}

\bibliography{bibliography}

@article{goodman1961snowball,
    ISSN = {00034851},
    URL = {http://www.jstor.org/stable/2237615},
    author = {Leo A. Goodman},
    journal = {The Annals of Mathematical Statistics},
    number = {1},
    pages = {148--170},
    publisher = {Institute of Mathematical Statistics},
    title = {Snowball Sampling},
    urldate = {2026-04-28},
    volume = {32},
    year = {1961}
}

@article{frank1977survey,
    author = {Frank, Ove},
    title = {Survey sampling in graphs},
    journal = {Journal of Statistical Planning and Inference},
    volume = {1},
    pages = {235-264},
    year = {1977}
}

@article{frank1994estimating,
    author = {Frank, Ove and Snijders, Tom A. B.},
    title = {Estimating the Size of Hidden Populations Using Snowball Sampling},
    journal = {Journal of Official Statistics},
    year = {1994},
    pages = {53-57},
    volume = {10 (1)}
}

@article{snijders1992estimation,
    author = {Snijders, Tom A. B.},
    title = {Estimation on the Basis of Snowball Samples: How To Weight?},
    journal = {BMS: Bulletin of Sociological Methodology},
    year = {1992},
    pages = {59-70},
    volume = {36}
}

@article{vincent2022estimating,
    author = {Vincent, Kyle and Thompson, Steve},
    title = {Estimating the Size and Distribution of Networked Populations with Snowball Sampling},
    journal = {Journal of Survey Statistics and Methodology},
    volume = {10 (2)},
    pages = {397–418},
    year = {2022}
}

@article{robins2007intro,
    title = {An introduction to exponential random graph (p*) models for social networks},
    journal = {Social Networks},
    volume = {29},
    number = {2},
    pages = {173-191},
    year = {2007},
    note = {Special Section: Advances in Exponential Random Graph (p*) Models},
    issn = {0378-8733},
    doi = {https://doi.org/10.1016/j.socnet.2006.08.002},
    url = {https://www.sciencedirect.com/science/article/pii/S0378873306000372},
    author = {Garry Robins and Pip Pattison and Yuval Kalish and Dean Lusher}
}

@article{hoff2002latent,
    author = {Peter D Hoff and Adrian E Raftery and Mark S Handcock},
    title = {Latent Space Approaches to Social Network Analysis},
    journal = {Journal of the American Statistical Association},
    volume = {97},
    number = {460},
    pages = {1090--1098},
    year = {2002},
    publisher = {ASA Website},
    doi = {10.1198/016214502388618906}
}

@inbook{kaur2023latent,
    title = {The Sage Handbook of Social Network Analysis},
    publisher = {SAGE Publications Limited},
    author = {Kaur, Hardeep and Rastelli, Riccardo and Friel, Nial and Raftery, Adrian E.},
    year = {2023},
    pages = {526–541},
    chapter = {36},
}

@inbook{newman2018networks,
    title = {Networks: An Introduction},
    author = {Newman, Mark},
    chapter = {12},
    pages = {397-427},
    year = {2018},
    publisher = {Oxford University Press}
}

@article{smith2019geometry,
    title = {The Geometry of Continuous Latent Space Models for Network Data},
    author = {Smith, Anna L and Asta, Dena M and Calder, Catherine A},
    journal = {Statistical science},
    volume = {34},
    number = {3},
    pages = {428--453},
    year = {2019},
    publisher = {Institute of Mathematical Statistics}
}

@article{rastelli2016properties,
    title = {Properties of latent variable network models},
    author = {Rastelli, Riccardo and Friel, Nial and Raftery, Adrian E},
    journal = {Network Science},
    volume = {4},
    number = {4},
    pages = {407--432},
    year = {2016}
}

@book{lemieux2008quasi,
  author = {Lemieux, Christiane},
  title = {Monte Carlo and Quasi-Monte Carlo Sampling},
  year = {2008},
  publisher = {Springer}
}

@techreport{diebolt1995stochastic,
  title = {A stochastic EM algorithm for approximating the maximum likelihood estimate},
  author = {Diebolt, J and Ip, E HS},
  year = {1995},
  institution = {Sandia National Lab.(SNL-CA), Livermore, CA (United States)}
}

@article{celeux1985sem,
  title = {The SEM algorithm: a probabilistic teacher algorithm derived from the EM algorithm for the mixture problem},
  author = {Celeux, Gilles},
  journal = {Computational statistics quarterly},
  volume = {2},
  pages = {73--82},
  year = {1985}
}

@article{liu1989limited,
  title = {On the limited memory BFGS method for large scale optimization},
  author = {Liu, Dong C and Nocedal, Jorge},
  journal = {Mathematical programming},
  volume = {45},
  number = {1},
  pages = {503--528},
  year = {1989},
  publisher = {Springer}
}

@article{nielsen2000stochastic,
    ISSN = {13507265},
    URL = {http://www.jstor.org/stable/3318671},
    author = {Søren Feodor Nielsen},
    journal = {Bernoulli},
    number = {3},
    pages = {457--489},
    publisher = {International Statistical Institute (ISI) and Bernoulli Society for Mathematical Statistics and Probability},
    title = {The Stochastic EM Algorithm: Estimation and Asymptotic Results},
    urldate = {2026-01-13},
    volume = {6},
    year = {2000}
}

@inproceedings{young2007random,
  title = {Random dot product graph models for social networks},
  author = {Young, Stephen J and Scheinerman, Edward R},
  booktitle = {International workshop on algorithms and models for the web-graph},
  pages = {138--149},
  year = {2007},
  organization = {Springer}
}

@article{krioukov2010hyperbolic,
  title = {Hyperbolic geometry of complex networks},
  author = {Krioukov, Dmitri and Papadopoulos, Fragkiskos and Kitsak, Maksim and Vahdat, Amin and Bogun{\'a}, Mari{\'a}n},
  journal = {Physical Review E—Statistical, Nonlinear, and Soft Matter Physics},
  volume = {82},
  number = {3},
  pages = {036106},
  year = {2010},
  publisher = {APS}
}

@article{snijders1997estimation,
  title = {Estimation and prediction for stochastic blockmodels for graphs with latent block structure},
  author = {Snijders, Tom AB and Nowicki, Krzysztof},
  journal = {Journal of classification},
  volume = {14},
  number = {1},
  pages = {75--100},
  year = {1997},
  publisher = {Springer}
}

@article{choi2014co,
  title = {Co-clustering separately exchangeable network data},
  author = {Choi, David and Wolfe, Patrick J},
  journal = {The Annals of Statistics},
  volume = {42},
  number = {1},
  pages = {29--63},
  year = {2014}
}

@article{gao2015rate,
  title = {Rate-optimal graphon estimation},
  author = {Gao, Chao and Lu, Yu and Zhou, Harrison H},
  journal = {The Annals of Statistics},
  volume = {43},
  number = {6},
  pages = {2624--2652},
  year = {2015}
}

@article{sischka2025stochastic,
  title = {Stochastic block smooth graphon model},
  author = {Sischka, Benjamin and Kauermann, G{\"o}ran},
  journal = {Journal of Computational and Graphical Statistics},
  volume = {34},
  number = {1},
  pages = {140--154},
  year = {2025},
  publisher = {Taylor \& Francis}
}

@article{illenberger2012estimating,
  title = {Estimating network properties from snowball sampled data},
  author = {Illenberger, Johannes and Fl{\"o}tter{\"o}d, Gunnar},
  journal = {Social Networks},
  volume = {34},
  number = {4},
  pages = {701--711},
  year = {2012},
  publisher = {Elsevier}
}

@article{fritz2023modelling,
    author = {Fritz, Cornelius and De Nicola, Giacomo and Kevork, Sevag and Harhoff, Dietmar and Kauermann, Göran},
    title = {Modelling the large and dynamically growing bipartite network of German patents and inventors},
    journal = {Journal of the Royal Statistical Society Series A: Statistics in Society},
    volume = {186},
    number = {3},
    pages = {557-576},
    year = {2023},
    month = {07},
    issn = {0964-1998},
    doi = {10.1093/jrsssa/qnad009},
    url = {https://doi.org/10.1093/jrsssa/qnad009},
    eprint = {https://academic.oup.com/jrsssa/article-pdf/186/3/557/50686247/qnad009.pdf},
}

@article{newman2001structure,
    author = {M. E. J. Newman },
    title = {The structure of scientific collaboration networks},
    journal = {Proceedings of the National Academy of Sciences},
    volume = {98},
    number = {2},
    pages = {404-409},
    year = {2001},
    doi = {10.1073/pnas.98.2.404},
    URL = {https://www.pnas.org/doi/abs/10.1073/pnas.98.2.404},
    eprint = {https://www.pnas.org/doi/pdf/10.1073/pnas.98.2.404}
 }

@article{hedges1992meta,
    author = {Larry V. Hedges},
    title = {Meta-Analysis},
    journal = {Journal of Educational Statistics},
    volume = {17},
    number = {4},
    pages = {279-296},
    year = {1992},
    doi = {10.3102/10769986017004279},
    URL = {https://doi.org/10.3102/10769986017004279},
    eprint = {https://doi.org/10.3102/10769986017004279},
}

@article{hartwig2020median,
    author = {Hartwig, Fernando P. and Davey Smith, George and Schmidt, Amand F. and Sterne, Jonathan A. C. and Higgins, Julian P. T. and Bowden, Jack},
    title = {The median and the mode as robust meta-analysis estimators in the presence of small-study effects and outliers},
    journal = {Research Synthesis Methods},
    volume = {11},
    number = {3},
    pages = {397-412},
    doi = {https://doi.org/10.1002/jrsm.1402},
    url = {https://onlinelibrary.wiley.com/doi/abs/10.1002/jrsm.1402},
    eprint = {https://onlinelibrary.wiley.com/doi/pdf/10.1002/jrsm.1402},
    year = {2020}
}

@article{wand1994multivariate,
    title = {Multivariate plug-in bandwidth selection},
    author = {Wand, MP and Jones, Chris},
    journal = {Computational Statistics},
    volume = {9},
    number = {2},
    pages = {97--116},
    year = {1994}
}

@article{duong2007ks,
    title = {ks: Kernel Density Estimation and Kernel Discriminant Analysis for Multivariate Data in R},
    volume = {21},
    url = {https://www.jstatsoft.org/index.php/jss/article/view/v021i07},
    doi = {10.18637/jss.v021.i07},
    number = {7},
    journal = {Journal of Statistical Software},
    author = {Duong, Tarn},
    year = {2007},
    pages = {1–16}
}

@article{duong2003plugin,
    author = {Tarn Duong and Martin Hazelton},
    title = {Plug-in bandwidth matrices for bivariate kernel density estimation},
    journal = {Journal of Nonparametric Statistics},
    volume = {15},
    number = {1},
    pages = {17--30},
    year = {2003},
    publisher = {Taylor \& Francis},
    doi = {10.1080/10485250306039},
    URL = {https://doi.org/10.1080/10485250306039},
    eprint = {https://doi.org/10.1080/10485250306039}
}

@article{hunter2008gof,
    author = {David R Hunter and Steven M Goodreau and Mark S Handcock},
    title = {Goodness of Fit of Social Network Models},
    journal = {Journal of the American Statistical Association},
    volume = {103},
    number = {481},
    pages = {248--258},
    year = {2008},
    publisher = {Taylor \& Francis},
    doi = {10.1198/016214507000000446},
    URL = {https://doi.org/10.1198/016214507000000446},
    eprint = {https://doi.org/10.1198/016214507000000446}
}

@article{shore2015spectral,
    title = {Spectral goodness of fit for network models},
    journal = {Social Networks},
    volume = {43},
    pages = {16-27},
    year = {2015},
    issn = {0378-8733},
    doi = {https://doi.org/10.1016/j.socnet.2015.04.004},
    url = {https://www.sciencedirect.com/science/article/pii/S0378873315000301},
    author = {Jesse Shore and Benjamin Lubin}
}

@inbook{chung1997spectral,
    chapter = {1},
    title = {Spectral graph theory},
    author = {Chung, Fan RK},
    volume = {92},
    year = {1997},
    publisher = {American Mathematical Soc.},
    pages = {1-22}
}

@article{frank1986markov,
    author = {Ove Frank and David Strauss},
    title = {Markov Graphs},
    journal = {Journal of the American Statistical Association},
    volume = {81},
    number = {395},
    pages = {832--842},
    year = {1986},
    publisher = {Taylor \& Francis},
    doi = {10.1080/01621459.1986.10478342},
    URL = {https://www.tandfonline.com/doi/abs/10.1080/01621459.1986.10478342},
    eprint = {https://www.tandfonline.com/doi/pdf/10.1080/01621459.1986.10478342}
}

@misc{hu2013survey,
    title={A Survey and Taxonomy of Graph Sampling}, 
    author={Pili Hu and Wing Cheong Lau},
    year={2013},
    eprint={1308.5865},
    archivePrefix={arXiv},
    primaryClass={cs.SI},
    url={https://arxiv.org/abs/1308.5865} 
}

@article{zhang2017graph,
	title = {Graph sampling},
	volume = {75},
	issn = {2281-695X},
	url = {https://doi.org/10.1007/s40300-017-0126-y},
	doi = {10.1007/s40300-017-0126-y},
	pages = {277--299},
	number = {3},
	journal = {{METRON}},
	author = {Zhang, L.-C. and Patone, M.},
	year = {2017}
}

@article{heckathorn2017network,
   author = {Heckathorn, Douglas D. and Cameron, Christopher J.},
   title = {Network Sampling: From Snowball and Multiplicity to Respondent-Driven Sampling},
   journal= {Annual Review of Sociology},
   year = {2017},
   volume = {43},
   number = {43},
   pages = {101-119},
   doi = {https://doi.org/10.1146/annurev-soc-060116-053556},
   url = {https://www.annualreviews.org/content/journals/10.1146/annurev-soc-060116-053556},
   publisher = {Annual Reviews},
   issn = {1545-2115},
   type = {Journal Article},
}

@article{johnson1989estimating,
    title = {Estimating relational attributes from snowball samples through simulation},
    journal = {Social Networks},
    volume = {11},
    number = {2},
    pages = {135-158},
    year = {1989},
    issn = {0378-8733},
    doi = {https://doi.org/10.1016/0378-8733(89)90009-9},
    url = {https://www.sciencedirect.com/science/article/pii/0378873389900099},
    author = {J.C. Johnson and J.S. Boster and D. Holbert}
}

@misc{oguzalper2023snowball,
    title = {Snowball sampling from graphs}, 
    author = {Melike Oguz-Alper and Li-Chun Zhang},
    year = {2023},
    eprint = {2003.09467},
    archivePrefix = {arXiv},
    primaryClass = {stat.ME},
    url = {https://arxiv.org/abs/2003.09467}, 
}

@article{handcock2010modeling,
    title = {Modeling social networks from sampled data},
    author = {Handcock, Mark S and Gile, Krista J},
    journal = {The Annals of applied statistics},
    volume = {4},
    number = {1},
    pages = {5},
    year = {2010}
}

@article{thompson2000model,
    title = {Model-based estimation with link-tracing sampling designs},
    author = {Thompson, Steven K and Frank, Ove},
    journal = {Survey methodology},
    volume = {26},
    number = {1},
    pages = {87--98},
    year = {2000}
}

@article{pattison2013conditional,
    title = {Conditional estimation of exponential random graph models from snowball sampling designs},
    journal = {Journal of Mathematical Psychology},
    volume = {57},
    number = {6},
    pages = {284-296},
    year = {2013},
    note = {Social Networks},
    issn = {0022-2496},
    doi = {https://doi.org/10.1016/j.jmp.2013.05.004},
    url = {https://www.sciencedirect.com/science/article/pii/S0022249613000497},
    author = {Philippa E. Pattison and Garry L. Robins and Tom A.B. Snijders and Peng Wang}
}

@article{stivala2016snowball,
    title = {Snowball sampling for estimating exponential random graph models for large networks},
    journal = {Social Networks},
    volume = {47},
    pages = {167-188},
    year = {2016},
    issn = {0378-8733},
    doi = {https://doi.org/10.1016/j.socnet.2015.11.003},
    url = {https://www.sciencedirect.com/science/article/pii/S0378873315000878},
    author = {Alex D. Stivala and Johan H. Koskinen and David A. Rolls and Peng Wang and Garry L. Robins}
}

@article{holland1983stochastic,
    title = {Stochastic blockmodels: First steps},
    journal = {Social Networks},
    volume = {5},
    number = {2},
    pages = {109-137},
    year = {1983},
    issn = {0378-8733},
    doi = {https://doi.org/10.1016/0378-8733(83)90021-7},
    url = {https://www.sciencedirect.com/science/article/pii/0378873383900217},
    author = {Paul W. Holland and Kathryn Blackmond Laskey and Samuel Leinhardt}
}

@inbook{crane2018probabilistic,
    place = {Boca Raton, FL},
    title = {Probabilistic Foundations of Statistical Network Analysis},
    chapter = {3},
    pages = {25-50},
    publisher = {CRC Press},
    author = {Crane, Harry},
    year = {2018}
}

@article{ibrahim1990incomplete,
    author = {Joseph G. Ibrahim},
    title = {Incomplete Data in Generalized Linear Models},
    journal = {Journal of the American Statistical Association},
    volume = {85},
    number = {411},
    pages = {765--769},
    year = {1990},
    publisher = {Taylor \& Francis},
    doi = {10.1080/01621459.1990.10474938},
    URL = {https://www.tandfonline.com/doi/abs/10.1080/01621459.1990.10474938},
    eprint = {https://www.tandfonline.com/doi/pdf/10.1080/01621459.1990.10474938}
}

@InProceedings{murray2010elliptical,
  title = {Elliptical slice sampling},
  author = {Murray, Iain and Adams, Ryan and MacKay, David},
  booktitle = {Proceedings of the Thirteenth International Conference on Artificial Intelligence and Statistics},
  pages = {541--548},
  year = {2010},
  editor = {Teh, Yee Whye and Titterington, Mike},
  volume = {9},
  series = {Proceedings of Machine Learning Research},
  address = {Chia Laguna Resort, Sardinia, Italy},
  month = {13--15 May},
  publisher = {PMLR},
  url = {https://proceedings.mlr.press/v9/murray10a.html}
}

@article{watts1998collective,
  title = {Collective dynamics of ‘small-world’networks},
  author = {Watts, Duncan J and Strogatz, Steven H},
  journal = {Nature},
  volume = {393},
  number = {6684},
  pages = {440--442},
  year = {1998},
  publisher = {Nature Publishing Group}
}

@article{telesford2011ubiquity,
  title = {The ubiquity of small-world networks},
  author = {Telesford, Qawi K and Joyce, Karen E and Hayasaka, Satoru and Burdette, Jonathan H and Laurienti, Paul J},
  journal = {Brain connectivity},
  volume = {1},
  number = {5},
  pages = {367--375},
  year = {2011},
  publisher = {SAGE Publications Sage CA: Los Angeles, CA}
}

@article{lovasz2006limits,
  title = {Limits of dense graph sequences},
  journal = {Journal of Combinatorial Theory, Series B},
  volume = {96},
  number = {6},
  pages = {933-957},
  year = {2006},
  issn = {0095-8956},
  doi = {https://doi.org/10.1016/j.jctb.2006.05.002},
  url = {https://www.sciencedirect.com/science/article/pii/S0095895606000517},
  author = {László Lovász and Balázs Szegedy}
}

@misc{sapargali2026exact,
  title = {{Exact Likelihood Inference for Snowball-Sampled Erd\H{o}s-R\'enyi Networks}}, 
  author = {Nurzhan Sapargali and Sergio Buttazzo and G\"oran Kauermann},
  year = {2026},
  eprint = {2608.14129},
  archivePrefix = {arXiv},
  primaryClass = {stat.ME},
  url = {https://arxiv.org/abs/2608.14129}, 
}

\newpage

\appendix

\section{Likelihood Derivation for CLS Networks}\label{app:cls_likelihood_derivation}

The likelihood of a snowball sample conditional on latent variables under the CLS population model can be derived by mirroring the derivation of the wave-set probabilities and the conditional probability of the sampled adjacency matrix under the ER model given by \citet{sapargali2026exact}, replacing the constant edge probability $\pi$ with the dyad-specific edge probabilities $\pi_\alpha(z_i, z_j)$.
Given the latent positions $Z = (z_1, \dots, z_N)$ of all vertices in the population and the initial wave set $V^{(0)} = \{i_0\}$, consisting of the ego vertex $i_0$, the chain probabilities of the wave sets $\V^{(1)}, \dots, \V^{(r)}$ can be expressed as
\begin{equation*}
    P(V^{(1)} \mid V^{(0)}, z) = \prod_{i \in V^{(1)}} \pi_\alpha(z_i, z_{i_0}) \prod_{j \in V\setminus (V^{(0)} \cup V^{(1)})} (1 - \pi_\alpha(z_j, z_{i_0})),
\end{equation*}
\begin{align*}
    P(V^{(k)} \mid V^{(0)}, \dots, V^{(k-1)}, z) &= \prod_{i \in V^{(k)}} \big[ 1 - \prod_{t \in V^{(k-1)}} (1 - \pi_\alpha(z_i, z_t)) \big] \\
    &\quad \cdot \prod_{j \in V \setminus \bigcup_{s = 0}^{k} V^{(s)}} \prod_{t \in V^{(k-1)}} (1 - \pi_\alpha(z_j, z_t)).
\end{align*}
Together, these enforce the recruitment half of the wave-consistency constraint defining $\mathcal{Y}$ (Section~\ref{subsec:er_recap}): a vertex enters wave $k$ only if it has at least one edge to wave $k-1$.
The complementary half, the absence of edges between non-adjacent waves, is enforced below.
For notational economy, we suppress the indicator $\mathds{1}\{\ry \in \mathcal{Y}\}$ throughout this derivation and reintroduce it only in the final expression.

Analogous to the derivation of \citet{sapargali2026exact}, for the conditional probability of the sampled adjacency matrix $\rY = \ry$ given the wave sets $V^{(0)}, \dots, V^{(r)}$ and the latent positions $Z$, we first partition the set of observed vertex pairs into the within-wave pairs $W$, between non-adjacent wave pairs $J$, and between adjacent wave pairs $A$.

The probabilities of the partitions $\rY_W$, $\rY_J$ of the sampled adjacency matrix are
\begin{gather*}
    P(\ry_{J} \mid V^{(0)}, \dots, V^{(r)}, z) = 1 \quad \text{for} \quad \ry_J = \boldsymbol{0}, \\
    P(\ry_{W} \mid \ry_{J}, V^{(0)}, \dots, V^{(r)}, z) = P(\ry_{W} \mid z) = \prod_{\{i,j\} \in W} \pi_\alpha(z_i, z_j)^{\ry_{i,j}} (1 - \pi_\alpha(z_i, z_j))^{1 - \ry_{i,j}}
\end{gather*}
For $\rY_A$, the set of adjacent wave pairs $A$ can again be partitioned into the disjoint sets $A(i)$ for all vertices in waves $1$ to $r$ as in \citet{sapargali2026exact}, where $A(i) = \{\{i,j\} \in A : j \in V^{(v(i) - 1)}\}$ is the set of vertex pairs in $A$ that include vertex $i$ and all the vertices in the previous wave $V^{(v(i) - 1)}$.
The conditional probability of $\rY_{A(i)} = \ry_{A(i)}$ given the wave sets and latent positions is
\begin{align*}
    P(\ry_{A(i)} \mid V^{(0)}, \dots, V^{(r)}, z) &= P\Big(\ry_{A(i)} \mid \sum_{\{u,l\} \in A(i)} \rY_{u,l} > 0, z \Big) \\
    &= \frac{P(\ry_A \mid z)}{P\Big(\sum_{\{u,l\} \in A(i)} \rY_{u,l} > 0 \mid z \Big)} = \frac{P(\ry_A \mid z)}{1 - P(\rY_{A(i)} = \boldsymbol{0} \mid z)} \\
    &= \frac{\prod_{\{u,l\} \in A(i)} \pi_\alpha(z_u, z_l)^{\ry_{u,l}}(1 - \pi_\alpha(z_u, z_l))^{1 - \ry_{u,l}}}{1 - \prod_{\{u,l\} \in A(i)} (1 - \pi_\alpha(z_u, z_l))}.
\end{align*}

Similar to the ER model, exploiting the independence of the sets $A(i)$ for all $i \in \bigcup_{k = 1}^r V^{(k)}$ yields the conditional probability of $\rY_A = \ry_A$ given the wave sets and latent positions
\begin{align*}
    P(\ry_A \mid V^{(0)}, \dots, V^{(r)}, z) &= P\Big(\ry_A \mid \sum_{\{u,l\} \in A(i)} \rY_{u,l} > 0 \text{ for all } i \in \bigcup_{k = 1}^r V^{(k)}, z \Big) \\
    &= \prod_{i \in \bigcup_{k = 1}^r V^{(k)}} P\Big(\ry_{A(i)} \mid \sum_{\{u,l\} \in A(i)} \rY_{u,l} > 0, z \Big) \\
    &= \prod_{i \in \bigcup_{k = 1}^r V^{(k)}} \frac{\prod_{\{u,l\} \in A(i)} \pi_\alpha(z_u, z_l)^{\ry_{u,l}}(1 - \pi_\alpha(z_u, z_l))^{1 - \ry_{u,l}}}{1 - \prod_{\{u,l\} \in A(i)} (1 - \pi_\alpha(z_u, z_l))} \\
    &= \frac{\prod_{\{i,j\} \in A} \pi_\alpha(z_i, z_j)^{\ry_{i,j}}(1 - \pi_\alpha(z_i, z_j))^{1 - \ry_{i,j}}}{\prod_{i \in \bigcup_{k = 1}^r V^{(k)}} \left[1 - \prod_{\{u,l\} \in A(i)} (1 - \pi_\alpha(z_u, z_l)) \right]} \\
    &= \frac{\prod_{\{i,j\} \in A} \pi_\alpha(z_i, z_j)^{\ry_{i,j}}(1 - \pi_\alpha(z_i, z_j))^{1 - \ry_{i,j}}}{\prod_{k = 1}^r \prod_{i \in V^{(k)}} \left[1 - \prod_{j \in V^{(k-1)}} (1 - \pi_\alpha(z_i, z_j)) \right]},
\end{align*}
where the last equality follows from the fact that $A(i)$ consists of the vertex pairs $\{i,j\}$ for all $j \in V^{(k-1)}$ if $i \in V^{(k)}$.

Thus, the likelihood of the sampled adjacency matrix $\rY = \ry$ given the wave sets $V^{(0)}, \dots, V^{(r)}$ and the latent positions $Z$ can be expressed as
\begin{align*}
    P(\ry \mid V^{(0)}, \dots, V^{(r)}, z) &= P(\ry_A \mid V^{(0)}, \dots, V^{(r)}, z)P(\ry_W \mid z) \\
    &= \frac{\prod_{\{i,j\} \in A} \pi_\alpha(z_i, z_j)^{\ry_{i,j}}(1 - \pi_\alpha(z_i, z_j))^{1 - \ry_{i,j}}}{\prod_{k = 1}^r \prod_{i \in V^{(k)}} \left[1 - \prod_{j \in V^{(k-1)}} (1 - \pi_\alpha(z_i, z_j)) \right]} \\
    &\quad \cdot \prod_{\{i,j\} \in W} \pi_\alpha(z_i, z_j)^{\ry_{i,j}}(1 - \pi_\alpha(z_i, z_j))^{1 - \ry_{i,j}} \\
    &= \frac{\prod_{\{i,j\} \in A \cup W} \pi_\alpha(z_i, z_j)^{\ry_{i,j}}(1 - \pi_\alpha(z_i, z_j))^{1 - \ry_{i,j}}}{\prod_{k = 1}^r \prod_{i \in V^{(k)}} \left[1 - \prod_{j \in V^{(k-1)}} (1 - \pi_\alpha(z_i, z_j)) \right]}.
\end{align*}

Bringing together the likelihood of the sampled adjacency matrix given the wave sets and latent positions with the chain probabilities of the wave sets yields the likelihood of a snowball sample conditional on latent variables under CLS population:
\begin{align*}
    P(\ry, V^{(1)}, \dots, &V^{(r)} \mid V^{(0)}, z) = P(\ry \mid V^{(0)}, \dots, V^{(r)}, z) \prod_{k = 1}^r P(V^{(k)} \mid V^{(0)}, \dots, V^{(k-1)}, z) \\
    &= \frac{\prod_{\{i,j\} \in A \cup W} \pi_\alpha(z_i, z_j)^{\ry_{i,j}}(1 - \pi_\alpha(z_i, z_j))^{1 - \ry_{i,j}}}{\prod_{k = 1}^r \prod_{i \in V^{(k)}} \left[1 - \prod_{j \in V^{(k-1)}} (1 - \pi_\alpha(z_i, z_j)) \right]} \\
    &\quad \cdot \prod_{i \in V^{(1)}} \pi_\alpha(z_i, z_{i_0}) \prod_{j \in V\setminus (V^{(0)} \cup V^{(1)})} (1 - \pi_\alpha(z_j, z_{i_0}))\\
    &\quad \cdot \prod_{k = 2}^r  \prod_{i \in V^{(k)}} \big[ 1 - \prod_{t \in V^{(k-1)}} (1 - \pi_\alpha(z_i, z_t)) \big] \cdot \prod_{j \in V \setminus \bigcup_{s = 0}^{k} V^{(s)}} \prod_{t \in V^{(k-1)}} (1 - \pi_\alpha(z_j, z_t)) \\
    &= \frac{\prod_{\{i,j\} \in A \cup W} \pi_\alpha(z_i, z_j)^{\ry_{i,j}}(1 - \pi_\alpha(z_i, z_j))^{1 - \ry_{i,j}}}{\prod_{i \in V^{(1)}} \left[1 - \prod_{j \in V^{(0)}} (1 - \pi_\alpha(z_i, z_j)) \right]} \\
    &\quad \cdot \prod_{i \in V^{(1)}} \pi_\alpha(z_i, z_{i_0}) \prod_{j \in V\setminus (V^{(0)} \cup V^{(1)})} (1 - \pi_\alpha(z_j, z_{i_0})) \\
    &\quad \cdot \prod_{k = 2}^r \prod_{j \in V \setminus \bigcup_{s = 0}^{k} V^{(s)}} \prod_{t \in V^{(k-1)}} (1 - \pi_\alpha(z_j, z_t)) \\
\end{align*}
Because $V^{(0)} = \{i_0\}$, the denominator can be simplified to $\prod_{i \in V^{(1)}} \pi_\alpha(z_i, z_{i_0})$, which cancels the corresponding term in the second term
\begin{align*}
    P(\ry, V^{(1)}, \dots, V^{(r)} \mid V^{(0)}, z) &= \prod_{\{i,j\} \in A \cup W} \pi_\alpha(z_i, z_j)^{\ry_{i,j}}(1 - \pi_\alpha(z_i, z_j))^{1 - \ry_{i,j}} \\
    &\quad \cdot \prod_{j \in V\setminus (V^{(0)} \cup V^{(1)})} (1 - \pi_\alpha(z_j, z_{i_0})) \\
    &\quad \cdot \prod_{k = 2}^r \prod_{j \in V \setminus \bigcup_{s = 0}^{k} V^{(s)}} \prod_{t \in V^{(k-1)}} (1 - \pi_\alpha(z_j, z_t)).
\end{align*}
As $\prod_{j \in V\setminus (V^{(0)} \cup V^{(1)})} (1 - \pi_\alpha(z_j, z_{i_0}))$ is equal to $\prod_{k = 1}^1 \prod_{j \in V \setminus \bigcup_{s = 0}^{k} V^{(s)}} \prod_{t \in V^{(k-1)}} (1 - \pi_\alpha(z_j, z_t))$, the second term can be absorbed into the product over $k$ simplifying the likelihood to
\begin{align*}
    P(\ry, V^{(1)}, \dots, V^{(r)} \mid V^{(0)}, z) &= \prod_{\{i,j\} \in A \cup W} \pi_\alpha(z_i, z_j)^{\ry_{i,j}}(1 - \pi_\alpha(z_i, z_j))^{1 - \ry_{i,j}} \\
    &\quad \cdot \prod_{k = 1}^r \prod_{i \in V \setminus \bigcup_{s = 0}^{k} V^{(s)}} \prod_{j \in V^{(k-1)}} (1 - \pi_\alpha(z_i, z_j)).
\end{align*}

Let $\mathcal{U} = V \setminus \bigcup_{s = 0}^r V^{(s)}$ denote the set of unsampled vertices.
We first observe that the complement of the cumulative sample at wave $k$ can be partitioned as:
\begin{equation*}
    V \setminus \bigcup_{s = 0}^k V^{(s)} = \begin{cases}
    \mathcal{U} \cup (\bigcup_{s = k + 1}^r V^{(s)}) & \text{for } k = 1, \dots, r - 1, \\
    \mathcal{U} & \text{for } k = r.
    \end{cases}
\end{equation*}
Using this partition, the product term above can be split into three distinct components:
\begin{align*}
    \prod_{k = 1}^{r} \prod_{i \in V \setminus \bigcup_{s = 0}^k V^{(s)}} \prod_{j \in V^{(k-1)}} (1 - \pi_\alpha(z_i,z_j)) &= \prod_{k = 1}^{r - 1} \prod_{i \in \bigcup_{s = k + 1}^r V^{(s)}} \prod_{j \in V^{(k-1)}} (1 - \pi_\alpha(z_i,z_j)) \\
    &\cdot \prod_{k = 1}^{r - 1} \prod_{i \in \mathcal{U}} \prod_{j \in V^{(k-1)}} (1 - \pi_\alpha(z_i,z_j)) \\
    &\cdot \prod_{i \in \mathcal{U}} \prod_{j \in V^{(r-1)}} (1 - \pi_\alpha(z_i,z_j)).
\end{align*}
Collecting the terms that involve the latent positions of unsampled vertices yields:
\begin{align*}
    \prod_{k = 1}^{r} \prod_{i \in V \setminus \bigcup_{s = 0}^k V^{(s)}} \prod_{j \in V^{(k-1)}} (1 - \pi_\alpha(z_i,z_j)) &= \prod_{k = 1}^{r - 1} \prod_{i \in \bigcup_{s = k + 1}^r V^{(s)}} \prod_{j \in V^{(k-1)}} (1 - \pi_\alpha(z_i,z_j)) \\
    &\cdot \prod_{i \in \mathcal{U}} \prod_{j \in \bigcup_{s = 0}^{r - 1} V^{(s)}} (1 - \pi_\alpha(z_i,z_j)).
\end{align*}
Substituting this expression back and reintroducing the indicator, we obtain a factorization of the likelihood into terms involving only sampled vertices and terms involving the unsampled set:
\begin{align*}
    P(\ry, V^{(1)}, \dots, V^{(r)} \mid V^{(0)}, z) &= \left[\prod_{\{i,j\} \in A \cup W} \pi_\alpha(z_i, z_j)^{\ry_{i,j}} (1 - \pi_\alpha(z_i,z_j))^{1 - \ry_{i,j}}\right] \\
    &\cdot \prod_{k = 1}^{r - 1} \prod_{i \in \bigcup_{s = k + 1}^r V^{(s)}} \prod_{j \in V^{(k-1)}} (1 - \pi_\alpha(z_i,z_j)) \\
    &\cdot \left[\prod_{i \in \mathcal{U}} \prod_{j \in \bigcup_{s = 0}^{r - 1} V^{(s)}} (1 - \pi_\alpha(z_i,z_j))\right] \mathds{1}\{\ry \in \mathcal{Y}\}. \numberthis\label{eq:lsm_likelihood_not_simplified}
\end{align*}
The second line of \eqref{eq:lsm_likelihood_not_simplified} involves vertices from non-adjacent waves, corresponding to the index set $J$ defined in Section~\ref{subsec:er_recap}.
The third line involves only pairs consisting of one unsampled vertex and one vertex from waves $0$ to $r-1$.
Observing that $\ry_{i,j} = 0$ for all $\{i,j\} \in J$ whenever $\ry \in \mathcal{Y}$, the full expression simplifies to
\begin{align*}
    P(\ry, V^{(1)}, \dots, V^{(r)} \mid V^{(0)}, z) &= \left[\prod_{\{i,j\} \in A \cup W \cup J} \pi_\alpha(z_i, z_j)^{\ry_{i,j}} (1 - \pi_\alpha(z_i,z_j))^{1 - \ry_{i,j}}\right] \\
    &\cdot \left[\prod_{i \in \mathcal{U}} \prod_{j \in \bigcup_{s = 0}^{r - 1} V^{(s)}} (1 - \pi_\alpha(z_i,z_j))\right] \mathds{1}\{\ry \in \mathcal{Y}\},
\end{align*}
which is precisely \eqref{eq:cls_joint_prob} of Section~\ref{subsec:cls_extension}.

\section{Example of Non-identifiability of Latent Positions of Unsampled Vertices}\label{app:lsm_counterexample}
Assume a CLS population network with parameter $\theta = (\alpha, \psi)$ as defined in Section~\ref{subsec:cls_model}.
An $r$-wave snowball sample is drawn from the population network under ignorable ego selection, so that the population prior distribution of the latent positions coincides with the distribution of the latent positions conditional on the ego vertex as discussed in Section~\ref{subsec:cls_extension}.
Consider the joint density of the observed data and latent positions, parameterized by $\theta = (\alpha, \psi)$.
From Section~\ref{subsec:lsm_estimation}, the complete data likelihood of the observed adjacency matrix $\rY = \ry$, wave sets $V^{(1)}, \dots, V^{(r)}$, and latent positions $z$ given the initial wave set $V^{(0)}$ can be expressed as
\begin{align*}
    f_\theta(\ry, V^{(1)}, \dots, V^{(r)}, z \mid V^{(0)}) =& \ P(\ry, V^{(1)}, \dots, V^{(r)} \mid V^{(0)}, z)\, f_\psi(z)  \\
    &= P(\ry, V^{(1)}, \dots, V^{(r)} \mid V^{(0)}, z)\, \prod_{i \in V} f_\psi(z_i)
\end{align*}
where $f(z_i)$ denotes the density of the latent position of vertex $i$ under the population prior distribution.
Following the likelihood result from Section~\ref{subsec:cls_extension} and isolating the terms that depend on the latent positions of unsampled vertices $z_{\mathcal{U}}$ yields
\begin{align*}
    f_\theta(\ry, V^{(1)}, \dots, V^{(r)}, z \mid V^{(0)}) =& \prod_{\{i,j\} \in A \cup W \cup J} \pi_\alpha(z_i, z_j)^{\ry_{i,j}} (1 - \pi_\alpha(z_i,z_j))^{1 - \ry_{i,j}} \\
    &\cdot \prod_{i \in \mathcal{U}} \prod_{j \in \bigcup_{s = 0}^{r - 1} V^{(s)}} (1 - \pi_\alpha(z_i,z_j)) \cdot \prod_{i \in V} f_\psi(z_i) \\
    =& h(z_{V \setminus \mathcal{U}}, \theta) \prod_{i \in \mathcal{U}} \prod_{j \in \bigcup_{s = 0}^{r - 1} V^{(s)}} (1 - \pi_\alpha(z_i,z_j)) \prod_{i \in \mathcal{U}} f_\psi(z_i)
\end{align*}
where $z_{V \setminus \mathcal{U}}$ denotes the latent positions of the sampled vertices and
\begin{equation*}
    h(z_{V \setminus \mathcal{U}}, \theta) = \prod_{\{i,j\} \in A \cup W \cup J} \pi_\alpha(z_i, z_j)^{\ry_{i,j}} (1 - \pi_\alpha(z_i,z_j))^{1 - \ry_{i,j}} \prod_{i \in V\setminus \mathcal{U}} f_\psi(z_i)
\end{equation*}
represents the collection of terms independent of the latent positions of unsampled vertices.

Consider the complete data likelihood evaluated at some parameter value $\tilde{\theta} = (\tilde{\alpha}, \tilde{\psi})$ and some configuration of latent positions $\tilde{z}$ yielding a likelihood value of $C$:
\begin{gather*}
    f_{\tilde{\theta}}(\ry, V^{(1)}, \dots, V^{(r)}, \tilde{z} \mid V^{(0)}) = C\\
    h(\tilde{z}_{V \setminus \mathcal{U}}, \tilde{\theta}) \prod_{i \in \mathcal{U}} \prod_{j \in \bigcup_{s = 0}^{r - 1} V^{(s)}} (1 - \pi_{\tilde{\alpha}}(\tilde{z}_i,\tilde{z}_j)) \prod_{i \in \mathcal{U}} f_{\tilde{\psi}}(\tilde{z}_i) = C \\
    \prod_{i \in \mathcal{U}} f_{\tilde{\psi}}(\tilde{z}_i) \prod_{j \in \bigcup_{k = 1}^r V^{(k-1)}} (1 - \pi_{\tilde{\alpha}}(\tilde{z}_i,\tilde{z}_j)) = \frac{C}{h(\tilde{z}_{V \setminus \mathcal{U}}, \tilde{\theta})} 
\end{gather*}

Take any $v \in \mathcal{U}$.
Suppose we change the configuration of latent positions by rescaling the positions of all other unsampled vertices $i \in \mathcal{U} \setminus \{v\}$ by $\lambda = (\lambda_i)_{i \in \mathcal{U} \setminus \{v\}}$.
We seek a scaling factor $\lambda_v$ for the remaining vertex $v$ such that the joint likelihood remains exactly $C$.
This requires solving with respect to $\lambda_v$
\begin{equation} \label{eq:lsm_identify_roots}
    f_{\tilde{\psi}}(\lambda_v \tilde{z}_v) \prod_{j \in \bigcup_{k = 1}^r V^{(k-1)}} (1 - \pi_{\tilde{\alpha}}(\lambda_v\tilde{z}_v,\tilde{z}_j)) = C_v
\end{equation}
where
\begin{equation*}
    C_v = \frac{C}{h(\tilde{z}_{V \setminus \mathcal{U}}, \tilde{\theta}) \prod_{i \in \mathcal{U} \setminus \{v\}} f_\psi(\lambda_i \tilde{z}_i) \prod_{j \in \bigcup_{k = 1}^r V^{(k-1)}} (1 - \pi_\alpha(\lambda_i \tilde{z}_i,\tilde{z}_j))}.
\end{equation*}

For a specific example, let the population network follow the distance model introduced in Section~\ref{sec:distance} with the Gaussian prior distribution of latent positions.
Since the Gaussian prior density decays to zero as $|\lambda_v| \to \infty$, the likelihood contribution of vertex $v$ approaches zero for large scalings.
By the Intermediate Value Theorem, a solution $\lambda_v$ exists provided that the maximum possible contribution of vertex $v$ exceeds the required residual likelihood $C_v$:
\begin{equation*}
    \sup_{\lambda_v \in \mathbb{R}} f_{\tilde{\psi}}(\lambda_v \tilde{z}_v) \prod_{j \in \bigcup_{k = 1}^r V^{(k-1)}} (1 - \pi_{\tilde{\alpha}}(\lambda_v\tilde{z}_v,\tilde{z}_j)) \geq C_v
\end{equation*}

\begin{figure}[t!]
    \centering
    \includegraphics[scale = 0.688]{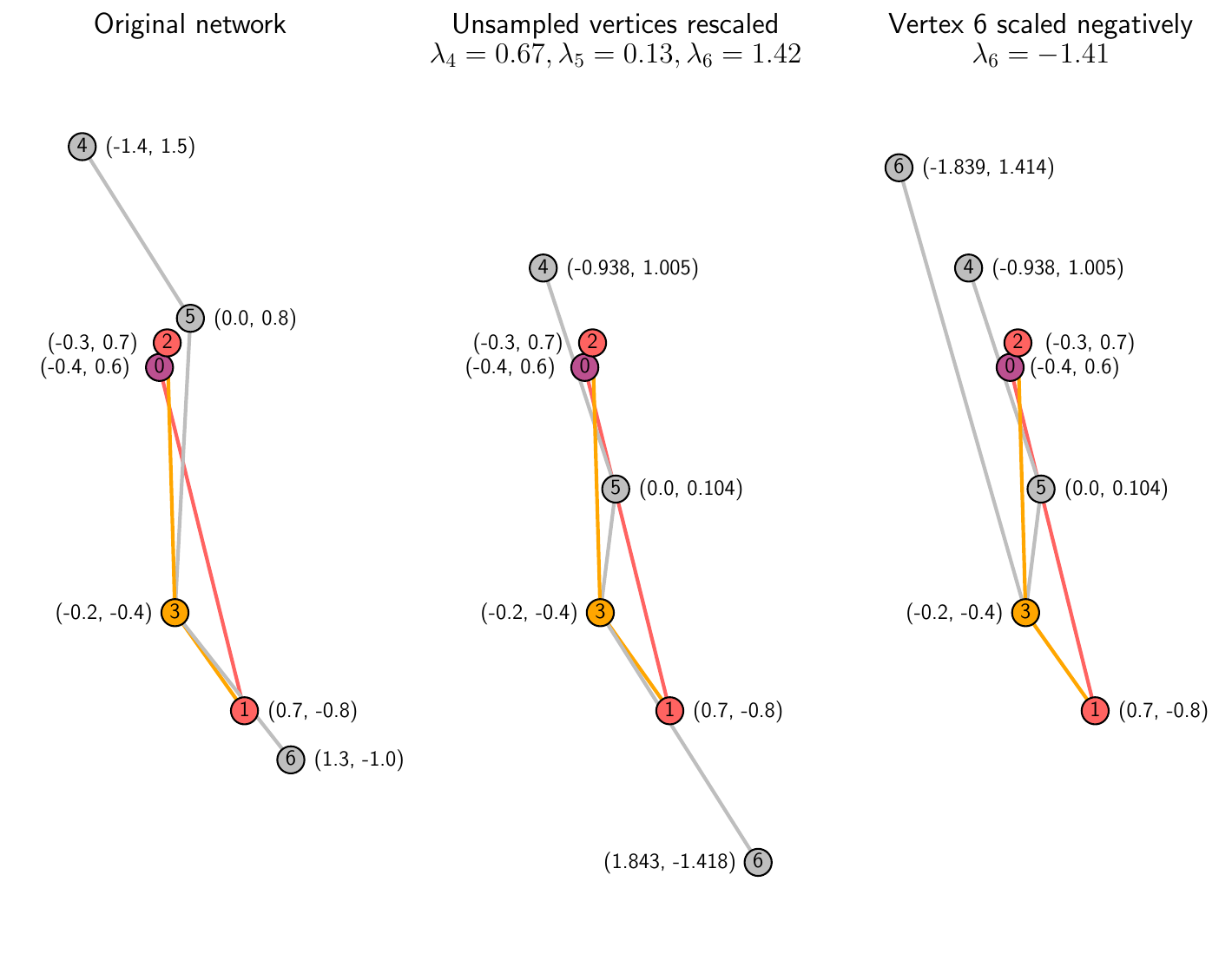}
    \caption{A 2-wave snowball sample originating from vertex 0 ($N=7$, $\alpha=0$, $\psi=0.5$).
    Red vertices constitute the first wave, the orange vertex constitutes the second wave, gray vertices ($4, 5, 6$) are unsampled.
    Vertices are placed in the latent space according to their positions in a two-dimensional latent space.
    Despite the visible shift in geometry, the alternative configurations (where unsampled positions are rescaled) yield an identical log-likelihood of $-32.39$.}
    \label{fig:identify_example}
\end{figure}

Figure~\ref{fig:identify_example} provides a numerical illustration.
In this example, we have a population of $N = 7$ vertices following the distance model with a two-dimensional latent space and parameters $\alpha = 0$ and $\psi = 0.5$.
A 2-wave snowball sample is drawn from vertex $0$ leaving vertices $4, 5,$ and $6$ unsampled.
We mechanically rescale the positions of vertices $4$ and $5$ by random factors $\lambda_4 = 0.67$ and $\lambda_5 = 0.13$ drawn from the uniform distribution on $(-1, 1.5)$.
The required scaling $\lambda_6$ for vertex $6$ is then computed numerically as the root of equation~\eqref{eq:lsm_identify_roots}.
This procedure yields two distinct geometric configurations for the unsampled vertices that are indistinguishable under the likelihood function.

\section{Stochastic Expectation-Maximization for Snowball-corrected Distance Model} \label{A:5}
\subsection{Stochastic E-step: Elliptical Slice Sampling within Gibbs}
\label{app:ess-gibbs}

The stochastic E-step draws latent positions $z_{V\setminus\mathcal{U}}$ from $f_{\hat\theta^{(t)}}(z_{V\setminus\mathcal{U}}\mid\ry,V^{(0)},\dots,V^{(r)})$ via a Gibbs sweep.
Vertices in $V\setminus\mathcal{U}$ are visited in sequence, and at each step $z_i$ is updated from its full conditional~\eqref{eq:full_cond} while all other positions $z_{-i}$ are held fixed.
Because the prior is Gaussian, each coordinate update uses Elliptical Slice Sampling (ESS) \citep{murray2010elliptical}, which is generally used to draw from a distribution over latent variables that is proportional to the product of a multivariate Gaussian prior and some likelihood function without tuning parameters.

Let $S = \bigcup_{k = 0}^{r - 1} V^{(k)}$ be the set of vertices that are not in the last wave and $\ell(z_i)$ denote the likelihood factor of the full conditional, i.e.\ the right-hand side of \eqref{eq:full_cond} with the prior $f_{\hat\psi^{(t)}}(z_i)$ omitted
\begin{align*}
    \log\ell(z_i) &= \sum_{j \in (\bigcup_{k=0}^r V^{(k)}) \setminus \{i\}} \Big[\ry_{ij}\log\pi_{\hat\alpha^{(t)}}(z_i,z_j) +(1-\ry_{ij})\log\big(1-\pi_{\hat\alpha^{(t)}}(z_i,z_j)\big) \Big] \\
    &\quad + \mathds{1}\!\bigl(i\notin V^{(r)}\bigr)\cdot|\mathcal{U}| \cdot\log\widehat{I}\big(\hat\theta^{(t)}, z_{S}\big), \numberthis \label{eq:ess_loglik}
\end{align*}
where the quasi-Monte Carlo (QMC) approximation of the expected exclusion probability~\eqref{eq:lsm_exclusion_prob} is
\begin{equation} \label{eq:qmc_approx}
    \widehat{I}\big(\hat\theta^{(t)}, z_{S}\big) = \frac{1}{2^M} \sum_{m=1}^{2^M} \prod_{j \in S} \big(1-\pi_\alpha(\sqrt\psi\,t_m,\,z_j)\big).
\end{equation}
with $t_1, \dots, t_{2^M} \sim N(0, I_d)$ generated via inverse transform of Sobol points.

A single coordinate update via ESS is described in Algorithm~\ref{alg:ess}.

The correction term in \eqref{eq:ess_loglik} vanishes for last-wave vertices $i\in V^{(r)}$.
Such vertices cannot attract unsampled vertices by construction, so their full conditional is unaffected by $\mathcal{U}$.
For a non-last-wave vertex $i\in S$, the position $z_i$ itself appears inside the product in \eqref{eq:qmc_approx}, so each ESS proposal $z'$ requires recomputing $\widehat{I}\big(\hat\theta^{(t)}, z_{S}\big)$.
This can be done efficiently by incrementally updating only the contribution of $i$ as described below.

\begin{algorithm}[h]
\caption{Elliptical Slice Sampling step for node $i$}
\label{alg:ess}
\begin{algorithmic}[1]
\Require Current position $z_i$; positions $z_{-i}$; parameters $\hat\theta^{(t)}=(\hat\alpha^{(t)},\hat\psi^{(t)})$; QMC points $t_1,\dots,t_{2^M}$
\State Draw $\nu \sim N(0,\hat\psi^{(t)}I_d)$
\State Draw $u \sim \mathrm{Uniform}(0,1)$;
  set $\ell^*\leftarrow\log\ell(z_i)+\log u$
\State Draw $w \sim \mathrm{Uniform}(0,2\pi)$;
  set $[w_{\min}, w_{\max}]\leftarrow[w - 2\pi, w]$
\Repeat
  \State $z'\leftarrow z_i \cos w + \nu \sin w$
  \If{$\log\ell(z')>\ell^*$}
    \State \Return{$z_i\leftarrow z'$}
  \EndIf
  \State \textbf{if} $w < 0$
    \textbf{then} $w_{\min} \leftarrow w$
    \textbf{else} $w_{\max} \leftarrow w$
  \State Draw $w \sim \mathrm{Uniform}(w_{\min}, w_{\max})$
\Until{$z'$ accepted}
\end{algorithmic}
\end{algorithm}

\paragraph{Incremental update of $\widehat{I}\big(\hat\theta^{(t)}, z_{S}\big)$.}
The approximation $\widehat{I}\big(\hat\theta^{(t)}, z_{S}\big)$ is an average over the transformed QMC points of a product over each non-last-wave vertex $j \in S$ of a probability of a non-edge. Define an $2^M \times |S|$ matrix where each entry is the logarithm of the associated non-edge probability for each QMC point and vertex $j$.
\begin{equation*}
  \mathbf{L} = (L_{m,j})_{1 \leq m \leq 2^M, j \in S}, \qquad  L_{m,j} = \log \big(1-\pi_{\hat\alpha^{(t)}}(\sqrt{\hat\psi^{(t)}} t_m, z_j)\big)
\end{equation*}
The row sum of $\mathbf{L}$ over the row $m$ is then $r_m = \sum_{j \in S} L_{m, j}$, so that 
\begin{equation} \label{eq:log_sum_qmc}
    \log \widehat{I}\big(\hat\theta^{(t)}, z_{S}\big) = -M\log 2 + \log \left(\sum_{m=1}^{2^M}e^{r_m}\right)
\end{equation}
where the last term on the right-hand side is computed via log-sum-exp trick to avoid numerical underflow
\begin{equation*}
    \log \left(\sum_{m=1}^{2^M}e^{r_m}\right) = r^* + \log \left(\sum_{m=1}^{2^M}e^{r_m - r^*} \right), \ \text{where} \ r^* = \max_{m = 1, \dots, 2^M} r_m .
\end{equation*}
When $z_i$ is updated to a proposal $z'$, the row sums are updated as $r'_m = r_m - L_{m,i} + L'_{m,i}$, where $L'_{m,i} = \log(1 - \pi_{\hat\alpha^{(t)}}(\sqrt{\hat\psi^{(t)}}t_m, z'))$.
Recomputing $\log \widehat{I}\big(\hat\theta^{(t)}, z_{S}\big)$ therefore costs $O(2^M)$ rather than $O(2^M S)$.
One Gibbs sweep consists of $|V\setminus\mathcal{U}|$ such ESS steps.
Throughout all simulations, we set $M = 9$.
For the real application, $M$ is decreased to $8$.

\subsection{M-step: gradient-based optimization of the $Q$-function}
\label{app:mstep}

The M-step maximizes the $Q$-function, defined as the log of the complete-data likelihood~\eqref{eq:complete_data_like} evaluated at $z^{(t)}_{V\setminus\mathcal{U}}$ and with the intractable expected exclusion probability replaced by its QMC approximation~\eqref{eq:qmc_approx}, resulting in the decomposition
\begin{equation*}
    Q(\theta; z^{(t)}_{V\setminus\mathcal{U}}) = Q_1(\alpha) + Q_2(\theta) + Q_3(\psi),
\end{equation*}
where
\begin{align*}
Q_1(\alpha) &= \sum_{\{i,j\}\in A\cup W\cup J} \big[\ry_{ij}\log\pi_\alpha(z^{(t)}_i,z^{(t)}_j) + (1-\ry_{ij})\log(1 -\pi_\alpha(z^{(t)}_i,z^{(t)}_j))\big], \\
Q_2(\theta) &= |\mathcal{U}|\log\widehat{I}(\theta, z^{(t)}_{S}), \\
Q_3(\psi) &= -\frac{|V\setminus\mathcal{U}|\,d}{2}\log(2\pi\psi) - \frac{1}{2\psi}\sum_{i\in V\setminus\mathcal{U}}\|z^{(t)}_i\|^2.
\end{align*}
We optimize over $(\alpha,\lambda)$ with $\lambda=\log\psi$, which removes the positivity constraint on $\psi$.
The analytical gradient $(\partial Q/\partial\alpha, \partial Q/\partial\lambda)$ is derived term by term below.

\paragraph{Gradient of $Q_1$.}
Differentiating the log-likelihood of observed dyads,
\begin{equation} \label{eq:dQ1}
\frac{\partial Q_1(\alpha)}{\partial\alpha} = \sum_{\{i,j\}\in A\cup W\cup J} [\ry_{ij}-\pi_\alpha(z^{(t)}_i,z^{(t)}_j)].
\end{equation}
$Q_1$ does not depend on $\lambda$.

\paragraph{Gradient of $Q_3$.}
Using $\psi=e^\lambda$,
\begin{equation} \label{eq:dQ3}
\frac{\partial Q_3(\psi)}{\partial\lambda} = -\frac{|V\setminus\mathcal{U}|\,d}{2} + \frac{1}{2\psi}\sum_{i\in V\setminus\mathcal{U}}\|z^{(t)}_i\|^2.
\end{equation}
$Q_3$ does not depend on $\alpha$.

\paragraph{Gradient of $Q_2$.}
Using the log-sum-exp representation \eqref{eq:log_sum_qmc} and letting $r_m = \sum_{j \in S} \log \big(1-\pi_{\alpha}(\sqrt{\psi} t_m, z^{(t)}_j)\big)$
\begin{align*}
    \frac{\partial}{\partial \alpha} \log\widehat{I}(\theta, z^{(t)}_S) &= \frac{\partial}{\partial \alpha} \log \left( \sum_{m = 1}^{2^M} e^{r_m} \right) = \left(\sum_{m = 1}^{2^M} e^{r_m}\right)^{-1} \sum_{m = 1}^{2^M} e^{r_m} \frac{\partial}{\partial \alpha} r_m \\
    &= \left(\sum_{m = 1}^{2^M} e^{r_m}\right)^{-1} \sum_{m = 1}^{2^M} e^{r_m} \sum_{j \in S} \frac{\partial}{\partial \alpha} \log(1 - \pi_\alpha(\sqrt{\psi} t_m, z^{(t)}_j)) \\
    &= -\left(\sum_{m = 1}^{2^M} e^{r_m}\right)^{-1} \sum_{m = 1}^{2^M} e^{r_m} \sum_{j \in S} \pi_\alpha(\sqrt{\psi} t_m, z^{(t)}_j)
\end{align*}
Thus,
\begin{equation} \label{eq:dQ2_alpha}
    \frac{\partial Q_2(\theta)}{\partial \alpha} = -|\mathcal{U}| \left(\sum_{m = 1}^{2^M} e^{r_m}\right)^{-1} \sum_{m = 1}^{2^M} e^{r_m} \sum_{j \in S} \pi_\alpha(\sqrt{\psi} t_m, z^{(t)}_j)
\end{equation}
For $\lambda$, similar derivation leads to
\begin{align*}
    \frac{\partial}{\partial \lambda} \log\widehat{I}(\theta, z^{(t)}_S) &= \left(\sum_{m = 1}^{2^M} e^{r_m}\right)^{-1} \sum_{m = 1}^{2^M} e^{r_m} \sum_{j \in S} \pi_\alpha(\sqrt{\psi} t_m, z^{(t)}_j) \frac{\partial}{\partial \lambda} \|\sqrt{\psi} t_m - z^{(t)}_j\| \\
    &= \left(\sum_{m = 1}^{2^M} e^{r_m}\right)^{-1} \sum_{m = 1}^{2^M} e^{r_m} \sum_{j \in S} \pi_\alpha(\sqrt{\psi} t_m, z^{(t)}_j) \frac{(\sqrt{\psi} t_m - z^{(t)}_j)^\top \sqrt{\psi} t_m}{2\|\sqrt{\psi} t_m - z^{(t)}_j\|},
\end{align*}
which yields
\begin{equation} \label{eq:dQ2_lambda}
    \frac{\partial Q_2(\theta)}{\partial \lambda} = |\mathcal{U}| \left(\sum_{m = 1}^{2^M} e^{r_m}\right)^{-1} \sum_{m = 1}^{2^M} e^{r_m} \sum_{j \in S} \pi_\alpha(\sqrt{\psi} t_m, z^{(t)}_j) \frac{(\sqrt{\psi} t_m - z^{(t)}_j)^\top \sqrt{\psi} t_m}{2\|\sqrt{\psi} t_m - z^{(t)}_j\|}
\end{equation}
The full gradient $(\partial Q/\partial\alpha,\;\partial Q/\partial\lambda)$ is obtained by summing \eqref{eq:dQ1}--\eqref{eq:dQ2_lambda}.
The M-step solves
\begin{equation} \label{eq:M_step}
    \hat\theta^{(t)} = \operatorname*{argmax}_{\alpha,\,\psi} Q(\theta; z^{(t)}_{V\setminus\mathcal{U}})
\end{equation}
using L-BFGS \citep{liu1989limited}, initialized at $(\hat\alpha^{(t-1)},\log\hat\psi^{(t-1)})$.

\subsection{Stochastic EM algorithm}
\label{app:sem}

Algorithm~\ref{alg:sem} gives the complete procedure.
The first $T_{\mathrm{burn}}$ iterations form a heat-up phase used to move the chain away from the initialization.
Each E-step is initialized at the final draw of the previous iteration
(warm start), avoiding a fresh burn-in at each step.

\begin{algorithm}[ht]
\caption{Stochastic EM for the snowball-corrected distance model}
\label{alg:sem}
\begin{algorithmic}[1]
\Require Initial parameters $\hat\theta^{(0)}=(\hat\alpha^{(0)},\hat\psi^{(0)})$;
  initial positions $z^{(0)}_{V\setminus\mathcal{U}}$;
  heat-up length $T_{\mathrm{burn}}$; Gibbs sweep count $g$;
  maximum number of iterations $T_{\max}$;
  QMC points $t_1,\dots,t_{2^M}$; tolerance $\tau$
\For{$t=1,2,\dots$}
  \State \textbf{Stochastic E-step}: run $g$ Gibbs sweeps (Algorithm~\ref{alg:ess}) from $z^{(t-1)}_{V\setminus\mathcal{U}}$ at $\hat\theta^{(t-1)}$;
    set $z^{(t)}_{V\setminus\mathcal{U}}$ to the final sweep's draw
  \State \textbf{M-step}: $\hat\theta^{(t)}\leftarrow \operatorname{argmax}_{\theta} Q(\theta; z^{(t)}_{V\setminus\mathcal{U}})$ via \eqref{eq:M_step}
  \If{($t>T_{\mathrm{burn}}$ \textbf{and} \eqref{eq:conv} holds) \textbf{or} $t = T_{\max}$}
    \State \textbf{break}
  \EndIf
\EndFor
\Ensure $\hat\theta^{(t)}$, $z^{(t)}_{V\setminus\mathcal{U}}$
\end{algorithmic}
\end{algorithm}

\paragraph{Convergence criterion.}
Let $\bar\alpha_t=t^{-1}\sum_{s=1}^{t}\hat\alpha^{(s)}$ and
$\bar\psi_t=t^{-1}\sum_{s=1}^{t}\hat\psi^{(s)}$ denote the running means of
the parameter iterates.
After the heat-up phase, the algorithm is declared converged when
\begin{equation} \label{eq:conv}
    \frac{|\bar\alpha_t-\bar\alpha_{t-1}|}{|\bar\alpha_{t-1}|} < \tau \qquad\text{and}\qquad \frac{|\bar\psi_t-\bar\psi_{t-1}|}{|\bar\psi_{t-1}|} < \tau
\end{equation}
hold for $30$ consecutive iterations.
We use $\tau = 10^{-3}$ throughout all simulations.
For the real application, we set $\tau = 0$, letting Algorithm~\ref{alg:sem} run for the maximum number of iterations.

\paragraph{Adaptive sweep schedule.}
The number of Gibbs sweeps $g$ per E-step is set as a decreasing function of the observed sample size $n_{\mathrm{obs}} = |V\setminus\mathcal{U}|$.
The schedule serves two purposes.
First, it keeps total estimation runtime manageable.
Since each Gibbs sweep evaluates pairwise interactions among all $n_{\mathrm{obs}}$ nodes, the per-sweep cost scales as $O(n_{\mathrm{obs}}^2)$, making large networks substantially more expensive per sweep.
More importantly, it matches sweep count to posterior concentration.
Small networks have diffuse posteriors and require more sweeps for adequate mixing, while large networks have concentrated posteriors and mix faster.
Concretely, with anchors $n_{\min}=100$, $n_{\max}=900$, $g_{\min}=100$, $g_{\max}=1000$, and power $p=2.5$, we set
\begin{equation*}
  g(n_{\mathrm{obs}}) =
  \begin{cases}
    g_{\max} & \text{if } n_{\mathrm{obs}} \leq n_{\min}, \\[4pt]
    g_{\min} + (g_{\max}-g_{\min})
      \left(1 - \dfrac{n_{\mathrm{obs}}-n_{\min}}{n_{\max}-n_{\min}}\right)^{\!p}
      & \text{if } n_{\min} < n_{\mathrm{obs}} < n_{\max}, \\[8pt]
    g_{\min} & \text{if } n_{\mathrm{obs}} \geq n_{\max}.
  \end{cases}
\end{equation*}
The exponent $p > 1$ makes the schedule decrease more steeply near $n_{\min}$, reflecting that even moderate increases in sample size from a small baseline substantially reduce the mixing time.

\section{Descriptive statistics of the semiconductor patent co-inventorship network}\label{A:6}
\begin{figure}[htbp]
  \centering
  \begin{minipage}{0.48\textwidth}
    \centering
    \includegraphics[width = 0.75\linewidth]{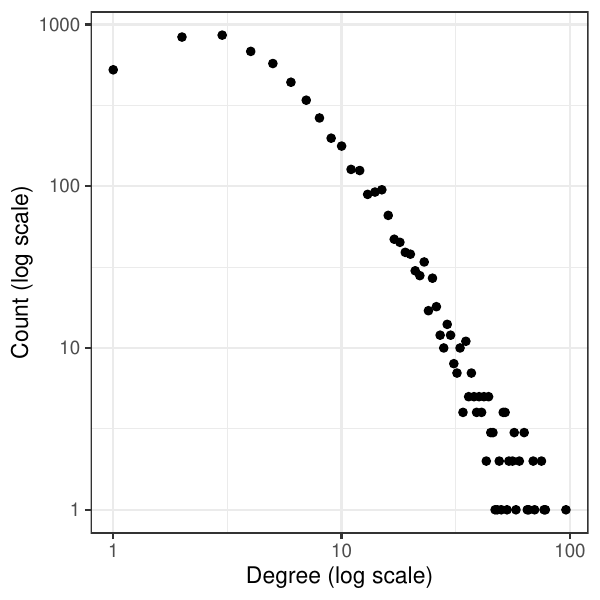}
    \caption{Degree distribution of the patent co-inventorship network in a log-log plot.}
    \label{fig:degree_dist}
  \end{minipage}
  \hfill
  \begin{minipage}{0.48\textwidth}
    \centering
    \begin{tabular}{lc}
        \hline
        Number of vertices & 5,979 \\
        Edge density & 0.0012  \\
        Average degree & 7.09 \\
        Standard deviation of degrees & 7.79 \\
        Maximum degree & 96 \\
        Diameter & 31 \\
        Average shortest path length & 7.55 \\
        Clustering coefficient & 0.31 \\
        Degree assortativity & 0.18 \\
        Clustering ratio to ER & 258.67 \\
        Path length ratio to ER & 1.70 \\
        Small-world index & 0.11 \\
        \hline
    \end{tabular}
    \captionof{table}{Descriptive statistics for the patent co-inventorship and citation networks.}
    \label{tab:descriptive}
  \end{minipage}
\end{figure}

\begin{figure}[htbp]
    \centering
    \includegraphics[width = \linewidth]{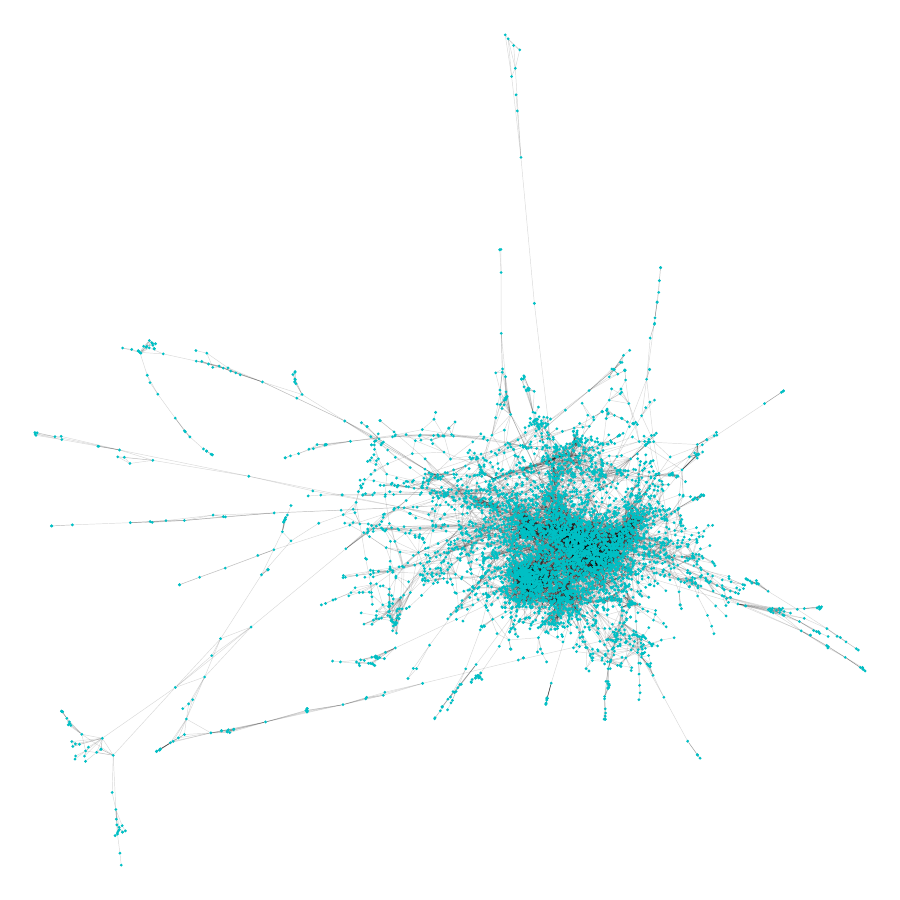}
    \caption{Co-inventorship of patent inventors in Germany over 1997-2012 in the area of semiconductors.}
    \label{fig:pop_network}
\end{figure}

Table \ref{tab:descriptive} provides descriptive statistics for the patent co-inventorship network, including the small-world index of \citet{telesford2011ubiquity}, while Figures~\ref{fig:pop_network} and \ref{fig:degree_dist} visualize the network and its degree distribution in a log-log plot, respectively.

\end{document}